\pdfoutput=1
\documentclass[cernpreprint,texlive=2016,LANGEDIT=true,UKenglish,texmf]{atlasdoc}

\usepackage{atlaspackage}
\usepackage{atlasbiblatex}
\usepackage{atlasphysics}
 
\usepackage{multirow}

\graphicspath{{logos/}{figures/}}
 
\usepackage{ANA-SUSY-2018-38-PAPER-defs}
\usepackage{graphicx}
\usepackage{amsmath}
 
\usepackage{siunitx}
\usepackage{booktabs}
\usepackage[version=4]{mhchem}
\usepackage{collcell}

\AtlasTitle{Search for phenomena beyond the Standard Model
in events with large $b$-jet multiplicity
using the ATLAS detector at the LHC}

\AtlasJournalRef{Eur. Phys. J. C 81 (2021) 11}
\AtlasDOI{10.1140/epjc/s10052-020-08730-0}
 
\PreprintIdNumber{CERN-EP-2020-147}
 
\AtlasAbstract{
A search is presented for new phenomena in
events characterised by high jet multiplicity, no leptons (electrons or muons), and four or more jets
originating from the fragmentation of $b$-quarks ($b$-jets). The search uses 139~\ifb\ of $\sqrt{s}$ = 13~\tev\ proton--proton collision data collected by the ATLAS experiment at the Large Hadron Collider during Run~2. The dominant
Standard Model background originates from multijet production and is estimated using a data-driven technique based on an extrapolation from events with low $b$-jet multiplicity to the
high $b$-jet multiplicities used in the search. No significant excess over the Standard Model expectation is observed and 95\% confidence-level limits that constrain  simplified models of R-parity-violating supersymmetry
are determined. The exclusion limits reach 950~\gev\ in top-squark mass in the models considered. \\}
 
\author{The ATLAS Collaboration}
 
\AtlasRefCode{SUSY-2018-38}

\AtlasHeading{ATLAS Paper Draft}

\hypersetup{pdftitle={ATLAS document},pdfauthor={The ATLAS Collaboration}}
 
\begin{document}
 
\maketitle

\section{Introduction}
\label{sec:intro}
Events with a large number of high-transverse momentum (\pt) jets originating from the fragmentation of $b$-quarks ($b$-jets) are rarely produced by Standard Model (SM) processes in proton--proton ($pp$) collisions at the LHC. 
As a result, this signature can provide sensitivity to certain
phenomena beyond the SM (BSM)~\cite{Evans:2014gfa,Buckley:2016kvr,Diglio:2016ynj}. Event signatures with five or more $b$-jets,
no leptons (electrons or muons) and without any requirements on missing transverse momentum
are not covered by existing searches at the LHC.

Supersymmetry (SUSY) provides an extension to the SM by introducing partners of the known bosons and fermions. It predicts the existence of superpartner
states (with different statistics) associated to each of the SM particles and fields. The lightest among such superpartners (LSP) may or may not be stable, depending on the conservation of   R-parity~\cite{Fayet:1974pd, Salam:1974xa,Barbier:2004ez}.
Final states with high leptonic or hadronic multiplicity are commonly predicted by R-parity-violating (RPV) SUSY. Models of RPV SUSY do not provide stable superpartners, and they give rise to a wide variety of experimental signatures whose nature depends on which of the many RPV couplings are non-zero.

\begin{figure}[htbp]
\centering
\subfloat[]{
\includegraphics[width=0.45\textwidth]{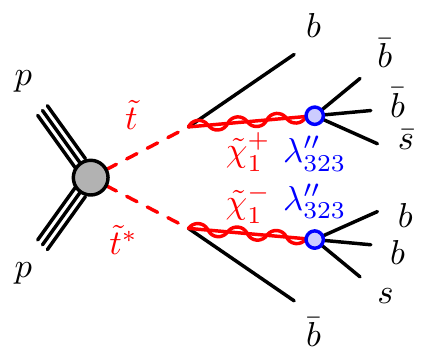}
}
\subfloat[]{
\includegraphics[width= 0.45\textwidth]{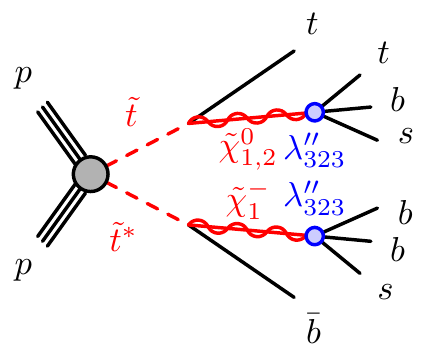}
}
\caption{ Diagrams of the 
signal processes involving pair production of top squarks $\tilde{t}$: (a) with the decay into a $b$-quark and the lightest chargino $ \tilde{\chi}^{+}_{1}$ ($ \tilde{t} \rightarrow b \tilde{\chi}^{+}_{1}$) with the subsequent decay of the chargino $\tilde{\chi}^{+}_1 \rightarrow \bar{b}\bar{b}\bar{s}$ and charge conjugate (c.c.), and (b) the decay into a top quark and the two lightest neutralinos $\tilde{\chi}^0_{1,2}$ with the subsequent decay $\tilde{\chi}^0_{1,2} \rightarrow tbs$.
}
\label{fig:diagram}
\end{figure}

In the analysis presented here, a particular benchmark model is considered in order to interpret the measurements in the
different jet and $b$-jet multiplicity regions. The process under consideration is the
pair production
of the top squark as the lightest of the coloured SUSY partners.
The existence of light SUSY partners of third-generation quarks, bottom squarks ($\tilde{b}$) and top squarks ($\tilde{t}$), is
favoured by naturalness considerations~\cite{Barbieri:1987fn,deCarlos:1993yy}.
The scenario assumes the LSP to be a triplet of two neutralino ($\tilde{\chi}^0_{1}$, $\tilde{\chi}^0_{2}$) and one chargino ($\tilde{\chi}^\pm_1$)
states that are mass-degenerate and carry dominantly higgsino components (in the following collectively referred to as ``higgsinos''). The top squark decays either into a chargino, $\tilde{\chi}^{\pm}_{1}$, and a bottom quark or into a neutralino, $\tilde{\chi}^{0}_{1,2}$, and a top quark. The chargino and neutralino decay, respectively, to $bbs$ and $tbs$ quark triplets, as shown in Figure~\ref{fig:diagram};
this decay is mediated
through their higgsino components via the non-zero baryon-number-violating RPV coupling $\lambda^{''}_{323}$~\cite{D_Ambrosio_2002,Cs_ki_2012}.

When $m_{\tilde{t}} - m_{\tilde{\chi}^0_{1,2},\tilde{\chi}^\pm_{1}} \leq \mtop$ (Figure~\ref{fig:diagram}(a)), the  $\tilde{t} \rightarrow t \tilde{\chi}^0_{1,2}$ decay is kinematically forbidden
and
the top-squark branching ratio ($B$) to $b \tilde{\chi}^\pm_1$ is equal to unity;
when $m_{\tilde{t}} - m_{\tilde{\chi}^0_{1,2},\tilde{\chi}^\pm_{1}} \geq \mtop$ the value of $B$
is taken to be 0.5.
In the latter case, the rest of the decay rate is evenly divided between the two neutralino states:
$\tilde{t} \rightarrow t \tilde{\chi}^0_{1,2} (\tilde{\chi}^0_{1,2} \rightarrow tbs)$ (Figure~\ref{fig:diagram}(b)). For the supersymmetric particle masses under consideration, the analysis considers only values of $\lambda^{''}_{323}\approx O(10^{-2}$--$10^{-1})$~\cite{Aaboud:2017faq} which ensure prompt neutralino and chargino decays and
omit more complex RPV decay patterns such as $\tilde{\chi}^\pm_1 \rightarrow  W^{\pm *} \tilde{\chi}^0_{1} (\tilde{\chi}^0_{1} \rightarrow tbs)$ or $\tilde{\chi}^0_{2} \rightarrow Z^* \tilde{\chi}^0_{1} (\tilde{\chi}^0_{1} \rightarrow tbs)$ that could be substantial for very small values of $\lambda^{''}_{323}$~\cite{Diglio:2016ynj}.

Previous searches targeting RPV SUSY models of pair-produced top squarks decaying through the coupling $\lambda^{''}_{323}$ have been carried out by the ATLAS and CMS collaborations. Those searches already exclude top-squark masses in the ranges 100~\GeV\  $\leq m_{\tilde{t}} \leq $ 470~\GeV\ and 480~\GeV\ $\leq m_{\tilde{t}} \leq $ 610~\GeV\ (ATLAS~\cite{Aaboud:2017nmi}), and 80~\GeV\  $\leq m_{\tilde{t}} \leq $ 270~\GeV, 285~\GeV\ $\leq m_{\tilde{t}} \leq $ 340~\GeV\ and 400~\GeV\ $\leq m_{\tilde{t}} \leq $ 505~\GeV\ (CMS~\cite{Sirunyan:2018rlj}) in scenarios where the top squark is the LSP and decays directly via $\tilde{t} \rightarrow bs$. For the direct top-squark production and $\lambda^{''}_{323}$-mediated decays of higgsino LSP scenarios, ATLAS has excluded top-squark masses up to 1.10~\TeV, depending on the higgsino mass considered, in the region where $m_{\tilde{t}} - m_{\tilde{\chi}^0_{1,2},\tilde{\chi}^\pm_1} \geq \mtop$, by analysing lepton plus jets events~\cite{Aaboud:2017faq}. CMS has excluded top-squark masses between 100 and 720~\GeV\ for
top-squark decays into four quarks in boosted topologies and with
the
mass
of the higgsinos set to 75\% of the squark mass~\cite{Sirunyan:2018zyx}.
 
This analysis considers events with six or more jets, of which at least four are identified as $b$-jets ($b$-tagged). There must be no identified electron or muon, and no requirement is made on the missing transverse momentum. In this channel, the dominant background is the non-resonant production of multijet events, referred to as `multijet' in the following, and a data-driven method is applied to estimate its yield.
Other backgrounds arise from top-quark pair production accompanied by extra $b$-jets or by a $Z$ or Higgs boson decaying into a $b$-quark pair. Results are
reported
as 95\% confidence level (CL) exclusion limits on the top-squark mass in the benchmark models described above. Model-independent limits on the possible contribution of BSM physics are also evaluated at large jet and $b$-tagged jet multiplicities.

\section{ATLAS detector}
\label{sec:detector}
\newcommand{\AtlasCoordFootnote}{
ATLAS uses a right-handed coordinate system with its origin at the nominal interaction point (IP)
in the centre of the detector and the \(z\)-axis along the beam pipe.
The \(x\)-axis points from the IP to the centre of the LHC ring,
and the \(y\)-axis points upwards.
Cylindrical coordinates \((r,\phi)\) are used in the transverse plane,
\(\phi\) being the azimuthal angle around the \(z\)-axis.
The pseudorapidity is defined in terms of the polar angle \(\theta\) as \(\eta = -\ln \tan(\theta/2)\).
Angular distance is measured in units of \(\Delta R \equiv \sqrt{(\Delta\eta)^{2} + (\Delta\phi)^{2}}\).}

The ATLAS experiment~\cite{PERF-2007-01} at the LHC is a multipurpose particle detector with a forward--backward symmetric cylindrical geometry and a near \(4\pi\) coverage in solid angle.\footnote{\AtlasCoordFootnote}
It consists of an inner tracking detector (ID) surrounded by a thin superconducting solenoid providing a \SI{2}{\tesla} axial magnetic field, electromagnetic and hadron calorimeters, and a muon spectrometer (MS).
The inner tracking detector covers the pseudorapidity range \(|\eta| < 2.5\). It consists of silicon pixel, silicon microstrip, and transition radiation tracking detectors.
An additional innermost layer of the silicon pixel tracker, the
insertable B-layer~\cite{ATLAS-TDR-2010-19,PIX-2018-001}, was installed in 2014 at an average radial distance of 3.3~cm from the beam-line
to improve track reconstruction and flavour identification of quark-initiated jets.
Lead/liquid-argon (LAr) sampling calorimeters provide electromagnetic energy measurements
with high granularity.
A steel/scintillator-tile calorimeter provides hadronic energy measurements and covers the central pseudorapidity range (\(|\eta| < 1.7\)).
The endcap and forward regions are instrumented with LAr calorimeters
for both the electromagnetic and hadronic energy measurements up to \(|\eta| = 4.9\).
The muon spectrometer surrounds the calorimeters and is based on
three large air-core toroidal superconducting magnets with eight coils each.
The field integral of the toroids ranges between \num{2.0} and \SI{6.0}{\tesla\metre}
across most of the detector.
The muon spectrometer includes a system of precision tracking chambers and fast detectors for triggering.
A two-level trigger system is used to select events to be recorded.
The first-level trigger is implemented in hardware and uses a subset of the detector information
to accept events at a rate of at most
\SI{100}{\kHz}.
This is followed by a software-based high-level trigger (HLT) that
reduces the accepted event rate to $\sim$\SI{1.2}{\kHz}, on average.
 
 
\section{Data collection and simulated event samples}
\label{sec:sim}
This search is based on 139 fb$^{-1}$ of centre-of-mass energy $\sqrt{s}=13$~\TeV\ $pp$ collision data,
collected between 2015 and 2018, that satisfy beam, detector and data-quality criteria. The uncertainty in the combined 2015--2018 integrated luminosity is 1.7\%~\cite{ATLAS-CONF-2019-021}, obtained using the LUCID-2 detector~\cite{Avoni_2018} for the primary luminosity measurements.
The average number of interactions ($\langle\mu\rangle$) in the same and nearby bunch crossings (pile-up) varies from $\langle\mu\rangle$ = 13.4 (2015 dataset) to $\langle\mu\rangle$ = 36.1 (2018 dataset), with a highest $\langle\mu\rangle$ = 37.8 (2017 dataset) and an average $\langle\mu\rangle$ = 33.7.
Data were collected using a four-jet trigger which, in the HLT, requires four jets each having $|\eta|<2.5$, with $\pt>100$~\GeV\ for the 2015--2016 data period and $\pt>120$~\GeV\ for the 2017--2018 data period.
Data events used for the validation of the data-driven multijet background were collected using the lowest unprescaled single-lepton triggers;
the lowest trigger \pt\ threshold used for muons is 20 (26)~\GeV\ in 2015 (2016--2018), while for electrons the trigger \pt\ threshold is 24 (26)~\GeV\ in 2015--2017 (2018).

Monte Carlo (MC) simulations are used to model the SUSY signals, as well as to aid in the description of the background processes.
In the remainder of this section, the simulation
of the signal and
of the main background processes contributing to the selected events in data
is described.
For all the simulated physics processes, the top-quark mass is assumed to be $\mtop = 172.5$~\gev\ and the Higgs boson mass is taken to be $\mH = 125$~\gev. The generation of the simulated event samples includes the effect of multiple $pp$ interactions in the same and neighbouring bunch crossings,
as well as the effect of pile-up on the detector response.
These interactions were produced using \textsc{Pythia~8.230}~\cite{Sjostrand:2014zea} with a set of tuned parameters called the A3 tune~\cite{ATL-PHYS-PUB-2016-017} and the NNPDF2.3 leading-order (LO)~\cite{Ball:2012cx} parton distribution function~(PDF) set.
 
All generated MC samples were processed through a simulation~\cite{Aad:2010ah} of the detector geometry and response using either
$\GEANT4$~\cite{Agostinelli:2002hh} or a fast simulation~\cite{Lukas:1458503} of the calorimeter response and were then
processed by the same reconstruction software used on data.
To model the parton shower, hadronisation, and underlying event,
the \textsc{Pythia 8} generator was used with the NNPDF2.3~LO PDF set and the A14~\cite{ATL-PHYS-PUB-2014-021} set of tunable parameters.
The decays of bottom and charm hadrons were
modelled using
\evtgen~\cite{EvtGen}. Simulated MC events are weighted such that the object identification efficiencies, energy scales and energy resolutions match those determined from data control samples~\cite{Aaboud:2017jcu,Aad:2019aic}.

MC samples for multijet production were generated using \textsc{Pythia~8.230} with leading-order matrix elements
for dijet production and
a \pt-ordered parton shower.
\evtgen~v1.6.0 was used for bottom and charm hadron decays.
The renormalisation and factorisation scales were set to the geometric
mean of the squared transverse masses of the two outgoing partons,
$\sqrt{(p_{\mathrm{T},1}^2 + m_1^2) (p_{\mathrm{T},2}^2 +
m_2^2)}$.
 
The production of $t\bar{t}$ events (referred to as \ttj) was modelled using the
\POWHEGBOX~v2~\cite{Frixione:2007nw,Nason:2004rx,Frixione:2007vw,Alioli:2010xd} generator at next-to-leading order (NLO) with the NNPDF3.0~NLO~\cite{Ball:2014uwa} PDF set and with the \hdamp\ parameter\footnote{The
\hdamp\ parameter is a resummation damping factor and one of the
parameters that controls the matching of \textsc{Powheg} matrix elements to
the parton shower and thus effectively regulates the
high-\pt\ radiation against which the \ttbar\ system recoils.} set
to 1.5~\mtop~\cite{ATL-PHYS-PUB-2016-020}.
\textsc{Pythia}~8.230 was used for the parton shower and \evtgen~v1.6.0 for
bottom and charm hadron decays.
The \ttj\ sample was generated inclusively in the number of jets using fast simulation.
The MC sample cross-section is corrected to the theory prediction at next-to-next-to leading order (NNLO) in QCD including resummation of next-to-next-to-leading logarithmic (NNLL) soft gluon terms by means of the \texttt{Top++} (v2.0) program~\cite{Beneke:2011mq,Cacciari:2011hy,Baernreuther:2012ws,Czakon:2012zr,Czakon:2012pz,Czakon:2013goa,Czakon:2011xx}.
The generated events may have jets which do not originate from the decay of the $\ttbar$ system. These additional jets are used to categorise the events depending on the flavour of the matching parton.
Particle jets are reconstructed from all stable particles generated in the event (excluding muons and neutrinos) using the anti-$k_t$ algorithm~\cite{Cacciari:2008gp} with a radius parameter $R=0.4$ and are required
to have $\pt>15\,\gev$ and $|\eta|<2.5$. Events having at least one such particle jet, matched within $\Delta R<0.3$ to a generated $b$-hadron having $\pt>5\,\gev$ and not originating from a top-quark decay, are labelled as \ttbin\ events. Similarly,
events which are not already categorised as \ttbin, and where at least one
particle jet is matched to a $c$-hadron not originating from a $W$ boson decay,
are labelled as \ttcin\ events.
Events labelled as either \ttbin\  or \ttcin\ are referred to as
\ttbar+HF events (HF for `heavy flavour').  The remaining events, including those with no additional jets, are labelled as $\ttbar$+light events (light for `light flavour').

The $Wt$  single-top-quark background was generated at NLO in QCD by
\POWHEGBOX~v2 with the NNPDF3.0~NLO PDF set.
Overlaps between the \ttbar\ and $Wt$ final states were
removed using the `diagram removal' scheme~\cite{Frixione:2008yi}.
\textsc{Pythia}~8.230 was used for the parton shower and \evtgen~v1.6.0 for
bottom and charm hadron decays.
Samples of single-top events are normalised to the cross-section calculated at NLO in QCD with NNLL soft gluon corrections~\cite{Kidonakis:2010ux,Kidonakis:2013zqa}.

The production of \ttV\ events was modelled using the
\MGMCatNLOV{v2.3.3}~\cite{Alwall:2014hca} generator at NLO with the
NNPDF3.0~NLO PDF set.
\textsc{Pythia}~8.210 was used for the parton shower and \evtgen~v1.2.0 for
bottom and charm hadron decays.

The production of \ttH\ events was modelled using the \POWHEGBOX~v2
generator to NLO with the NNPDF3.0~NLO PDF set.
\textsc{Pythia}~8.230 was used for the parton shower and \evtgen~v1.6.0 for
bottom and charm hadron decays.
The cross-sections are calculated at NLO QCD and NLO electroweak accuracy using \MGMCatNLO~\cite{deFlorian:2016spz}.

Signal events were produced using
the \MGMCatNLOV{v2.3.3} generator at NLO with the NNPDF2.3~LO PDF,
and the fast simulation of the detector response.
\textsc{Pythia}~8.230 was used for the parton shower and \evtgen~v1.6.0 for
bottom and charm hadron decays.
Signal cross-section calculations include approximate next-to-next-to-leading-order (NNLO$_{\textrm{Approx}}$) supersymmetric QCD corrections and the resummation of soft gluon emission
at NNLL accuracy~\cite{Beenakker:2016lwe}. 
The nominal cross-section and its uncertainty are taken from an envelope of predictions using different PDF sets as well as different factorisation and renormalisation scales. Top-squark masses between 600~\GeV\ and 1~\TeV\ and higgsino masses between 100~\GeV\ and 950~\GeV\ are considered.

\section{Event reconstruction}
\label{sec:event}
Events are required to have a primary vertex reconstructed from at least two tracks with transverse momentum \pt $>$ 500~\MeV. When several vertices are found in a given bunch crossing, the vertex with the largest summed $\pt^2$ of the associated tracks is selected as the primary vertex.
 
Electrons are reconstructed from energy deposits (clusters) in the electromagnetic calorimeter matched to tracks reconstructed in the ID~\cite{Aad_2014e, ATLAS-CONF-2016-024}
and are required to have $\pt>10$~\GeV\ and $|\eta|<2.47$. Candidates in the calorimeter barrel--endcap transition region ($1.37 <|\eta| < 1.52$) are excluded.
Electron tracks must match the primary vertex of the event: the longitudinal impact parameter\footnote{The
transverse impact parameter ($d_0$) is defined as the distance of closest approach in the transverse plane between a track and the beam-line. The longitudinal impact parameter ($z_0$) corresponds to the $z$-coordinate
difference
between the point along the track at which the transverse impact parameter is defined and the primary vertex.} is required to satisfy $|z_0|<0.5$~mm, while the transverse impact parameter is required to satisfy $|d_0|/\sigma_{d_0}<5$, where $\sigma_{d_0}$ represents the uncertainty in the measured $|d_0|$ values. \textit{Loose} electrons are identified using the `Medium' identification criterion provided by a likelihood-based discriminant~\cite{Aad_2019e}.
\textit{Tight} electrons are required to pass the `TightLH' selection~\cite{Aad_2019e} and the `Gradient' isolation criteria~\cite{Aad_2019e} and $\pt>27$~\GeV.

Muons are reconstructed by matching either track segments or full tracks in the MS to tracks in the ID~\cite{PERF-2015-10}.
Combined tracks are then re-fitted using information from both detector systems. Muon tracks must match the primary vertex of the event: the longitudinal impact parameter is required to satisfy $|z_0|<0.5$~mm, while the transverse impact parameter is required to satisfy $|d_0|/\sigma_{d_0}<3$.
\textit{Loose} muons are those that pass the `Loose' muon selection ~\cite{PERF-2015-10} and have $\pt>10$~\GeV\ and $|\eta|<2.5$, and \textit{Tight} muons are those that pass the `Medium' muon selection~\cite{PERF-2015-10}, satisfy the `FixedCutTightTrackOnly' isolation criterion~\cite{PERF-2015-10}, and have $\pt>27$~\GeV.
 
Jets are reconstructed from three-dimensional topological energy
clusters~\cite{Lampl:1099735} in the calorimeter using the anti-$k_{t}$ jet
algorithm~\cite{Cacciari:2008gp} with a radius parameter of 0.4.
Reconstructed jets are then corrected to the particle level by the application of a jet energy scale calibration that is derived from simulation and by \textit{in situ} corrections obtained from 13~\TeV\ data~\cite{Aad:2020flx}. Jets used in this analysis are required to have $\pt > 25~\gev$\ and $|\eta| < 2.5$ after calibration.
 
To avoid selecting jets from pile-up, low-\pt\ ($\pt < 120$~\GeV) jets in the central ($|\eta| < 2.5$) region of the
detector are required to satisfy the jet-vertex tagger~\cite{Aad:2015ina} configured such that it has an efficiency of approximately 92\% to identify jets from a primary vertex.
This requirement is applied to both data and simulation.
Quality criteria are imposed to identify jets arising from non-collision sources or detector noise (using the \texttt{BadLoose} operating
point~\cite{Run2-jet-cleaning}), and any event containing at least one such jet is removed. This  removal produces a negligible loss of efficiency for signal events.
 
The $b$-jets are identified via a $b$-tagging algorithm that uses multivariate techniques to combine information from the impact parameters of displaced tracks as well as topological properties of secondary and tertiary decay vertices reconstructed
within the jet. This analysis uses the MV2c10 tagger~\cite{ATL-PHYS-PUB-2016-012}, trained on a hybrid sample of simulated \ttbar\ and $Z'$ events statistically enriched at high-\pt\ in order to discriminate $b$-jets from a background consisting of light- ($93\%$)
and $c$-labelled ($7\%$) jets~\cite{Aad:2019aic}. A weight is calculated corresponding to the probable presence of a $b$-quark or a $c$-quark, and jets are confirmed $b$-tagged if they satisfy a minimum requirement on the MV2c10 $b$-tagging weight corresponding to an average efficiency in \ttbar events of 60\% for $b$-jets, 4\% for $c$-jets and a rejection factor of approximately 1200 for light-flavour jets across the jet \pt\ range.
 
An overlap removal procedure is carried out to resolve ambiguities between jets and lepton candidates.
To prevent treating electron energy deposits as jets, the closest jet within  $\Delta R_{y}=\sqrt{(\Delta y)^2+(\Delta \phi)^2}=0.2$
of a selected electron is removed.\footnote{The rapidity is defined as $y=\frac{1}{2} \ln\frac{E+p_{\text{z}}}{E-p_{\text{z}}}$
where $E$ is the energy and $p_{\text{z}}$ is the longitudinal component of the momentum along the beam-line.}
If the nearest jet surviving that selection is within $\Delta R_{y} = 0.4$ of the electron, the electron is discarded.
To reduce the background from heavy-flavour decays inside jets, muons are removed if they are separated from the nearest jet by $\Delta R_{y} < 0.4$. However, if that jet has fewer than three associated tracks, the muon is kept and the jet is removed instead.
\section{Analysis strategy}
\label{sec:analysis}
 
Events selected for further analysis
are required to have at least five jets, of which at least two must be $b$-tagged. The four highest-$\pt$ jets are required to be on the trigger efficiency plateau, namely to have $\pt > 120$~\GeV\ or $\pt > 140$~\GeV, depending on the jet-$\pt$ trigger requirement in 2015--2016 or 2017--2018, and have $|\eta|<2.5$. All other jets present in the event are required to have $\pt > 25$~\GeV\ and  $|\eta|<2.5$.
A lepton veto is applied: events
that contain \textit{loose} muons or electrons with $\pt > 10$~\GeV, whether isolated or non-isolated, are discarded.
 
After the selections described above, the largest background contribution to the measurement is from non-resonant multijet production from light-quark and gluonic final states. The next largest is from $t\bar{t}$+jets production. 
Other small background contributions originate from the production of a single top quark and from the production of a $t\bar{t}$ pair in association with either a vector boson or a Higgs boson. The estimation of the multijet background using a data-driven method
and the validation of this estimate
without significant
bias from potential signal contamination are the main challenges for this analysis.

\begin{table}[tbp]
\caption{
The strategy of the analysis.
For the model-dependent fit, the signal regions (SR$_{\tilde{t}}$) consist of events with $N_{\textrm{j}}=$ 6, 7, 8 and $\geq$ 9 jets and $N_{\textrm{b}}=$ 4 and $\geq$ 5. These are used independently in the final fit. For the model-independent fit, two dedicated signal regions (SR$_{\textrm{discovery}}$), with ($N_{\textrm{j}}\geq$ 9, $N_{\textrm{b}}\geq$ 5) and ($N_{\textrm{j}}\geq$ 8, $N_{\textrm{b}}\geq$ 5), are used. The validation regions (VR-MJ), which are based on a maximum value of the centrality mass, $C^{\textrm{max}}_{\textrm{mass}}$, introduced for the description of the VRs in Section~\ref{sec:analysis}, are also indicated.}
\begin{center}
\begin{tabular}{ c c l l l }
\toprule
\multicolumn{2}{l}{Analysis}  & \multicolumn{3}{c}{$N_{\textrm{b}}$} \\
\multicolumn{2}{l}{Regions}   &  \multicolumn{1}{c}{3} & \multicolumn{1}{c}{4}  & \multicolumn{1}{c}{$\geq5$} \\ [0.1cm]
\hline
\multirow{8}{*}{$N_{\textrm{j}}$} &  \multirow{2}{*}{6} & {\color{white} $\rule[0.5ex]{3em}{0.55pt}$ SR$_{\tilde{t}}$ $\rule[0.5ex]{3em}{0.55pt}$} & {\color{white} $\rule[0.5ex]{3em}{0.55pt}$ } SR$_{\tilde{t}}$ {\color{white} $\rule[0.5ex]{3em}{0.55pt}$ } & {\color{white} $\rule[0.5ex]{3em}{0.55pt}$ } SR$_{\tilde{t}}$ {\color{white} $\rule[0.5ex]{3em}{0.55pt}$ } \\   [0.1cm]
&                     & {\color{white} $\rule[0.5ex]{0.7em}{0.55pt}$ } VR-MJ ${\color{white}\rule[0.5ex]{0.2em}{0.55pt}}{{ C^{\textrm{max}}_{\textrm{mass}} = 1.2}}$ & {\color{white} $\rule[0.5ex]{0.7em}{0.55pt}$ } VR-MJ ${\color{white}\rule[0.5ex]{0.2em}{0.55pt}}{{ C^{\textrm{max}}_{\textrm{mass}} = 0.9}}$ & \\ [0.2cm] \cline{3-5}
&  \multirow{2}{*}{7} & & {\color{white} $\rule[0.5ex]{3em}{0.55pt}$ } SR$_{\tilde{t}}$ {\color{white} $\rule[0.5ex]{3em}{0.55pt}$ } & {\color{white} $\rule[0.5ex]{3em}{0.55pt}$ } SR$_{\tilde{t}}$ {\color{white} $\rule[0.5ex]{3em}{0.55pt}$ } \\   [0.1cm]
&                     & {\color{white} $\rule[0.5ex]{0.7em}{0.55pt}$ } VR-MJ ${\color{white}\rule[0.5ex]{0.2em}{0.55pt}}{{ C^{\textrm{max}}_{\textrm{mass}} = 1.2}}$ & {\color{white} $\rule[0.5ex]{0.7em}{0.55pt}$ } VR-MJ ${\color{white}\rule[0.5ex]{0.2em}{0.55pt}}{{ C^{\textrm{max}}_{\textrm{mass}} = 0.7}}$ & \\ [0.2cm] \cline{3-5}
&  \multirow{2}{*}{8} & & {\color{white} $\rule[0.5ex]{3em}{0.55pt}$ } SR$_{\tilde{t}}$ {\color{white} $\rule[0.5ex]{3em}{0.55pt}$ } & {\color{white} $\rule[0.5ex]{0.7em}{0.55pt}$ } SR$_{\tilde{t}}$, SR$_{\textrm{discovery}}$ {\color{white} $\rule[0.5ex]{0.7em}{0.55pt}$ } \\   [0.1cm]
&                     & {\color{white} $\rule[0.5ex]{0.7em}{0.55pt}$ } VR-MJ ${\color{white}\rule[0.5ex]{0.2em}{0.55pt}}{{ C^{\textrm{max}}_{\textrm{mass}} = 0.9}}$ & {\color{white} $\rule[0.5ex]{0.7em}{0.55pt}$ } VR-MJ ${\color{white}\rule[0.5ex]{0.2em}{0.55pt}}{{ C^{\textrm{max}}_{\textrm{mass}} = 0.5}}$ & \\ [0.2cm] \cline{3-5}
&  \multirow{2}{*}{$\geq$9} & & {\color{white} $\rule[0.5ex]{3em}{0.55pt}$ } SR$_{\tilde{t}}$ {\color{white} $\rule[0.5ex]{3em}{0.55pt}$ } & {\color{white} $\rule[0.5ex]{0.7em}{0.55pt}$ } SR$_{\tilde{t}}$, SR$_{\textrm{discovery}}$ {\color{white} $\rule[0.5ex]{0.7em}{0.55pt}$ } \\   [0.1cm]
&                     & {\color{white} $\rule[0.5ex]{0.7em}{0.55pt}$ } VR-MJ ${\color{white}\rule[0.5ex]{0.2em}{0.55pt}}{{ C^{\textrm{max}}_{\textrm{mass}} = 0.7}}$ & {\color{white} $\rule[0.5ex]{0.7em}{0.55pt}$ } VR-MJ ${\color{white}\rule[0.5ex]{0.2em}{0.55pt}}{{ C^{\textrm{max}}_{\textrm{mass}} = 0.4}}$ & \\ [0.2cm] 
 
\bottomrule
 
\end{tabular}
 
\end{center}
\label{tab:ana}
\end{table}

\begin{figure}[htb]
\centering
\subfloat[]{
\includegraphics[width= 0.49\textwidth]{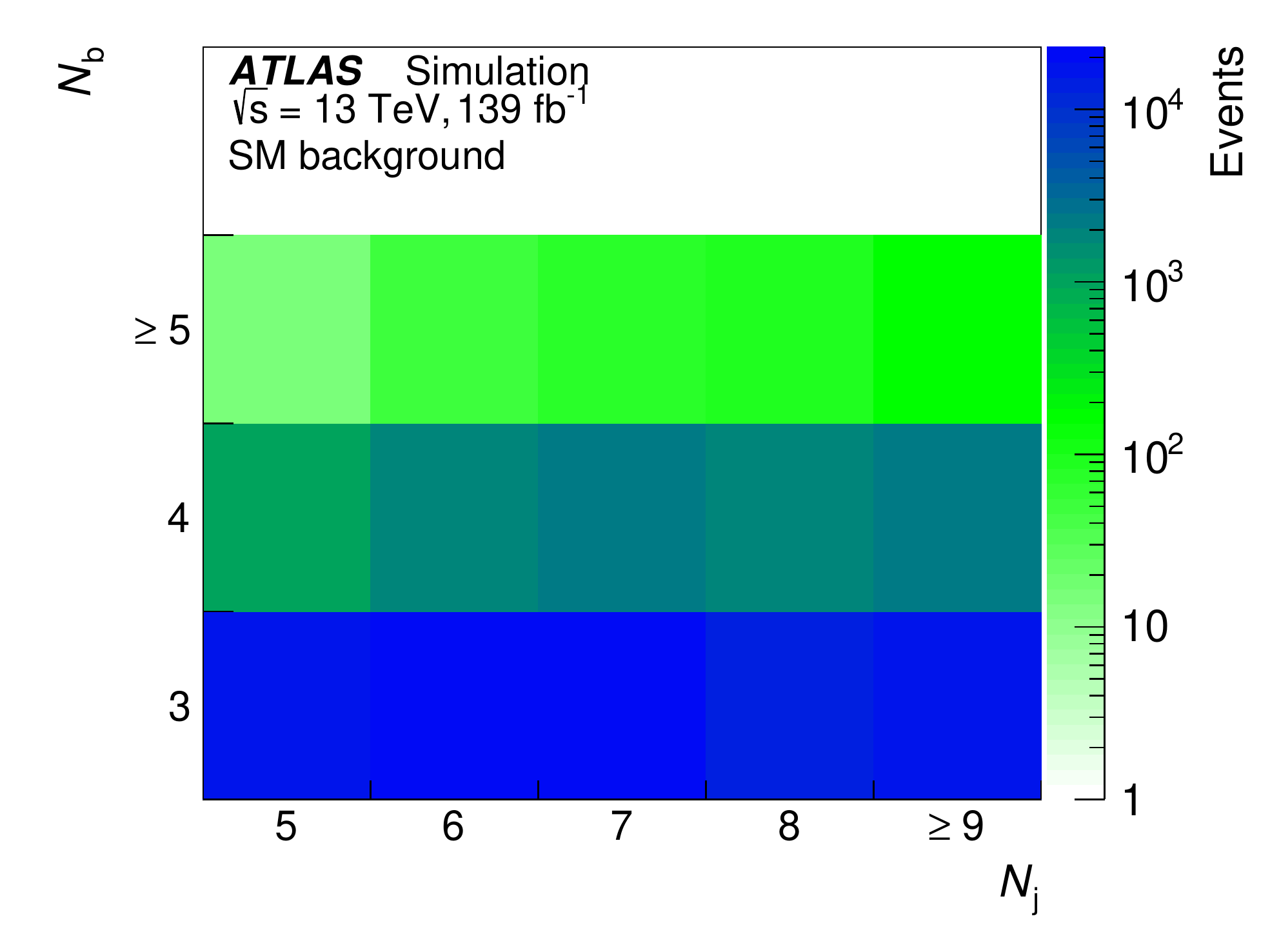}
}
\subfloat[]{
\includegraphics[width=0.49\textwidth]{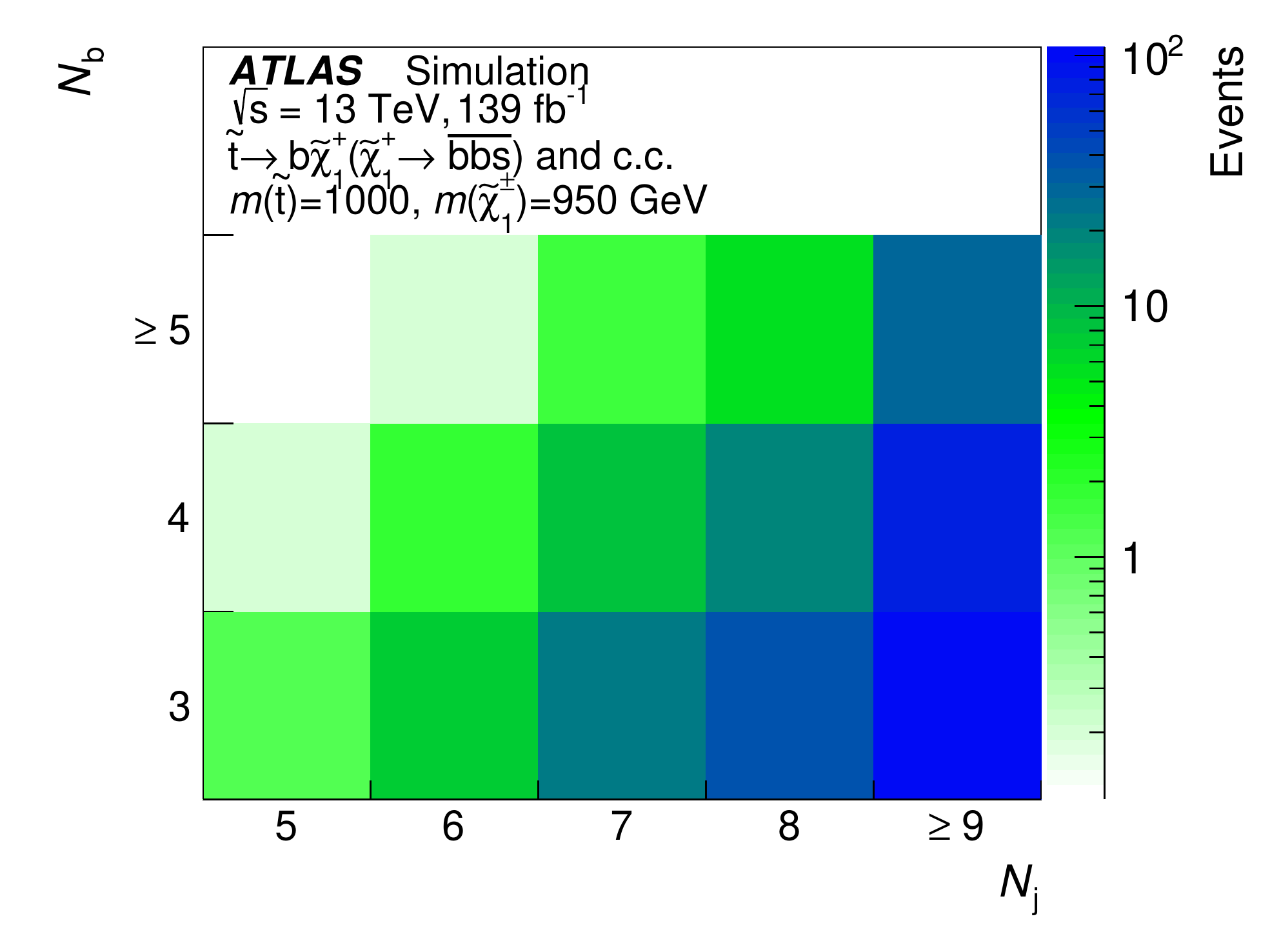}
}
\caption{Predicted numbers of events as a function of
jet multiplicity, $N_{\textrm{j}}$, and $b$-tagged jet multiplicity, $N_{\textrm{b}}$, for (a) SM background (multijet and top-quark production) and (b) top-squark pair production in the $\tilde{t} \rightarrow \bar{b} \tilde{\chi}^{+}_{1}$ ($\tilde{\chi}^{+}_{1} \rightarrow \bar{b}\bar{b}\bar{s}$) (and c.c.) channel, for $m_{\tilde{t}}=1000$~\GeV\ and $m_{\tilde{\chi}^{\pm}_{1}}=950$~\GeV. \label{fig:NjNb} }
\end{figure}

To probe top-squark pair production and estimate the contribution of signal top squarks in data, a model-dependent fit of the yield of events with jet multiplicity $N_{\textrm{j}}$ = 6, 7, 8 and $\geq$ 9 and $b$-tagged jet multiplicity $N_{\textrm{b}}$ = 4 and $\geq$ 5 is performed. These ($N_{\textrm{j}}$, $N_{\textrm{b}}$) regions are indicated as SR$_{\tilde{t}}$ in Table~\ref{tab:ana}. The signal contribution predicted for different values of $m_{\tilde{t}}$ and $m_{\tilde{\chi}^0_{1,2},\tilde{\chi}^{\pm}_{1}}$ is considered in all bins and is scaled by one common signal-strength parameter ($\mu_{\tilde{t}\tilde{t}^{*}}$). For the model considered here, the product of acceptance and reconstruction efficiency ($\textrm{A} \times \epsilon$) is of order $\sim5 \times 10^{-2}$ for $N_{\textrm{j}}\geq$ 9 and $N_{\textrm{b}}\geq$ 5. Figure~\ref{fig:NjNb} shows the number of signal events obtained from the model as a function of $N_{\textrm{j}}$ and $N_{\textrm{b}}$ compared to the estimated backgrounds. Their evaluation is described in Section~\ref{sec:mj}. The signal yields are concentrated at high jet and $b$-tagged jet multiplicity, while the backgrounds are concentrated at low $b$-tagged jet multiplicity. To validate the background estimates, intervals with $N_{\textrm{j}}=$ 6, 7, 8 and $\geq 9$, and $N_{\textrm{b}}=$ 3 and 4, subsequently referred to as VR-MJ, are used. In these, a region-dependent selection is applied, based on a maximum accepted value of the centrality mass ($C_{\textrm{mass}}$), defined as:
\begin{equation}\nonumber
C_{\textrm{mass}} = \frac{\htjet}{ \sqrt{ (\sum_{i=1}^{N_{\text{j}}} E_i)^2 - (\sum_{i=1}^{N_{\text{j}}} \textbf{p}_i)^2 }},
\end{equation}
i.e.\ the ratio of the scalar sum of all jet \pt\ in the event (\htjet) to the invariant mass of the set of observed jets. The signal-to-background ratio
decreases
monotonically with
decreasing $C_{\textrm{mass}}$
for all $N_{\textrm{j}}$ and $N_{\textrm{b}}$ values.
The value of the maximum value of $C_{\textrm{mass}}$ ($C^{\textrm{max}}_{\textrm{mass}}$) is chosen such that the signal-to-background ratio is less than 5\%.
Values of the $C^{\textrm{max}}_{\textrm{mass}}$ limits used are listed in Table~\ref{tab:ana}.

A separate, model-independent test is used to search for, and to set generic exclusion limits on, potential contributions from a hypothetical BSM signal by
comparing
the observed number of events
with
background predictions in two dedicated signal regions, one with $N_{\textrm{j}}\geq 9$ and $N_{\textrm{b}}\geq 5$ and the other with $N_{\textrm{j}}\geq 8$ and $N_{\textrm{b}}\geq 5$
(labelled SR$_{\textrm{discovery}}$ in Table~\ref{tab:ana}),
that were not
explored in previous searches at the LHC.
\section{Multijet background estimation}
\label{sec:mj}
The predominant multijet background is estimated via a data-driven method, subsequently referred to as  the tag-rate function method for multijet events (\trfmj)~\cite{Aad:2016zqi,Aaboud:2018jsj}. The aim is to extrapolate the $b$-tag multiplicity distributions from $N_{\textrm{j}}$ = 5, where the signal contamination for models not already excluded by other LHC searches is negligible, to larger $N_{\textrm{j}}$ values.
The \trfmj\ method uses a tag-rate function to quantify the experimental probability of $b$-tagging an additional jet in
samples of events with at least two,
or at least three, $b$-tagged jets.
This per-jet probability is then used to estimate the shape of the multijet $b$-tag multiplicity distribution for each $N_{\textrm{j}}$ value.
 
\begin{figure}[htbp]
\centering
\subfloat[]{
\includegraphics[width=0.49\textwidth]{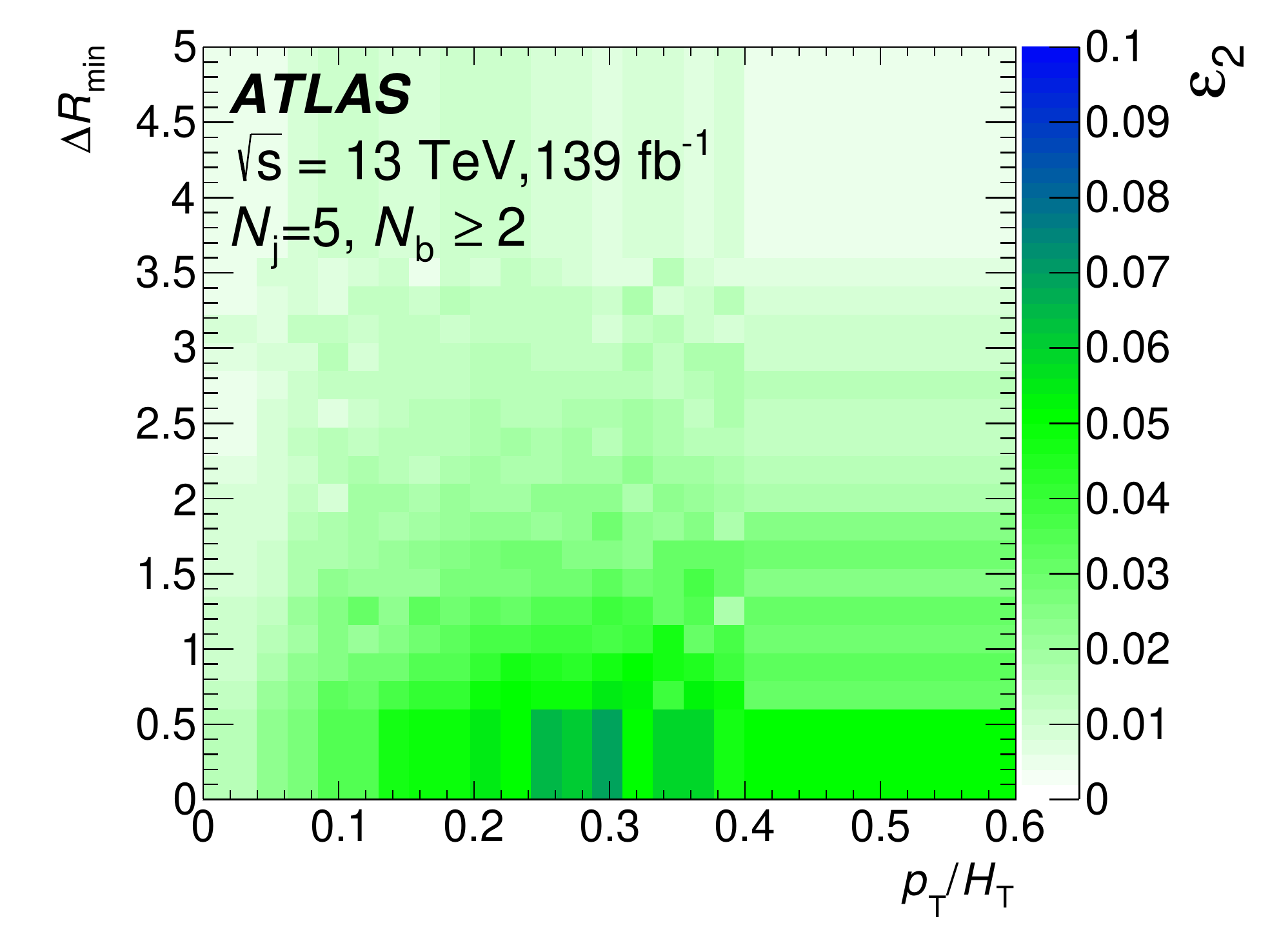}
\label{fig:mjeps2da}
}
\subfloat[]{
\includegraphics[width= 0.49\textwidth]{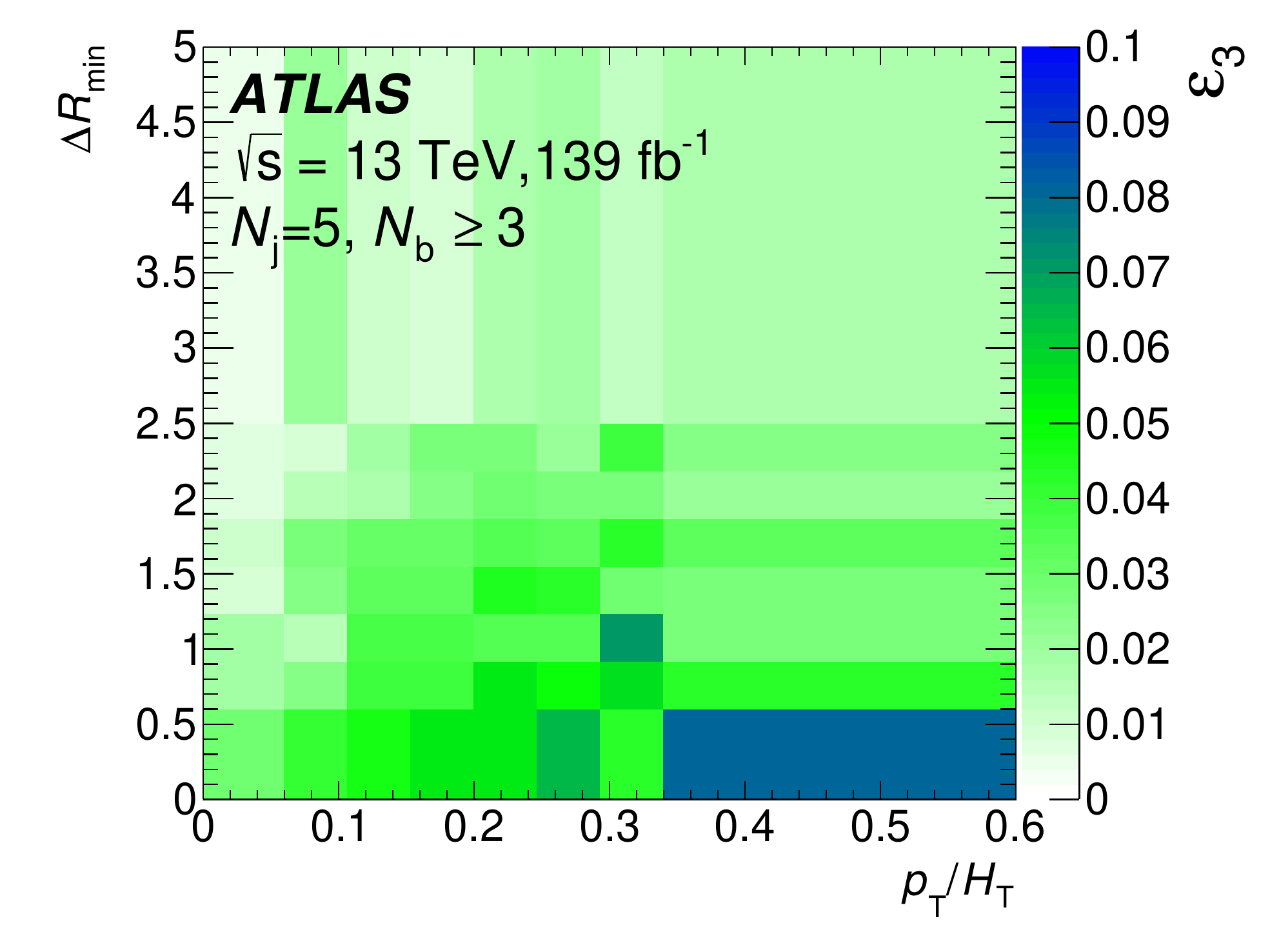}
\label{fig:mjeps2db}
}
\caption{Two-dimensional distributions of the probability (a) \mjepstwo\ or (b) \mjepsthree\ of $b$-tagging an additional jet in a sample of events with (a) at least two or (b) at least three $b$-tagged jets as a function of the ratio of jet transverse momentum to \HT, $p_{\text{T}}/H_{\text{T}}$, and the minimum $\Delta R$ between the jet and the (a) two or (b) three $b$-tagged jets with the highest $b$-tagging weight in the event, $\Delta R_{\text{min}}$. The choice of binning is made so as to avoid empty bins.}
\label{fig:mjeps2d}
\end{figure}

Events that satisfy the selection criteria described in Section~\ref{sec:analysis}
and that have exactly five jets, 
of which at least
two are $b$-tagged, are used to determine the $b$-tagging probability.
The data are first corrected by subtracting the expected non-multijet background found in simulation, approximately 5\% of the total.
After excluding the two jets in each event with the highest $b$-tagging weight, the probability that each remaining jet is $b$-tagged, denoted \mjepstwo, is calculated for this jet.  A similar procedure is used to calculate the probability \mjepsthree\ of additional $b$-tagged jets in events with at least three $b$-tagged jets.  These \mjeps\ probabilities are
parameterised as a function of both the \pt\ of the remaining jet
divided by \htjet, and the minimum $\Delta R$ between that jet and the two (for \mjepstwo)
or three (for \mjepsthree) jets with the largest $b$-tagging weight in the event ($\Delta R_{\text{min}}$).
This choice of variables for the
parameterisation is made to minimise the residual differences (non-closure) between the \trfmj\ prediction and the number of events obtained when selecting $b$-jets directly in the most sensitive signal regions in the multijet events simulated by MC.
The dependence
of \mjepstwo\ and \mjepsthree\ on both $\pt/\htjet$ and $\Delta R_{\text{min}}$
is shown in Figure~\ref{fig:mjeps2d}.
The rapid variation
with
$\Delta R_{\text{min}}$ is consistent with the dependence expected from multi-$b$-jet production due to gluon-splitting. The \pt/\htjet\ dependence, more visible at small $\Delta R_{\text{min}}$, reflects the variation of the $b$-tagging efficiency with jet \pt.
 
Following the methods of Ref.~\cite{Abazov:2006ka}, in the second step of the \trfmj\ method the expected number of events with each different number of $b$-tagged jets is estimated for each $N_{\textrm{j}}$ value by weighting all events with $N_{\textrm{b}} \geq 2$ by the
\textit{event probability}
of having $N_{\textrm{b}}=2$, 3, 4 and $\geq$5, respectively.  Upon subtracting the non-multijet background contribution~\cite{Aad:2016zqi}, the event probabilities are estimated using both \mjepstwo\ and \mjepsthree, after first excluding the two jets with the highest $b$-tagging weight.  For $N_{\textrm{b}}=2$ the event probabilities are estimated directly from \mjepstwo, treating the tagging probability for each jet as independent.
For 
$N_{\textrm{b}}=3$, 4 and $\geq$5, a two-step procedure is employed. First, a `pseudodata sample' with $N_{\textrm{b}}\geq3$ is emulated, using \mjepstwo\ in events with $N_{\textrm{b}} \geq 2$.
The additional emulated $b$-tagged jet is chosen randomly from the remaining $N_{\textrm{j}}-2$ jets by using their probability-dependent $b$-tagging weights~\cite{Aaboud:2018jsj}.
This emulated sample is then used
to estimate the event probabilities,
this time relying on \mjepsthree.
The probability of finding $N_{\textrm{b}}=4$ and $N_{\textrm{b}} \geq 5$ is estimated using the emulated $N_{\textrm{b}}\geq3$ sample via \mjepsthree. Due to too few events in the control sample from which
the \mjepstwo\ and \mjepsthree\ values are extracted, it is not possible to estimate the probability of $b$-tagging an additional jet in a sample of events with at least four $b$-tagged jets.

\subsection{Validation of \trfmj\ method}
 
\begin{figure}[h]
\centering
\includegraphics[width=0.75\textwidth]{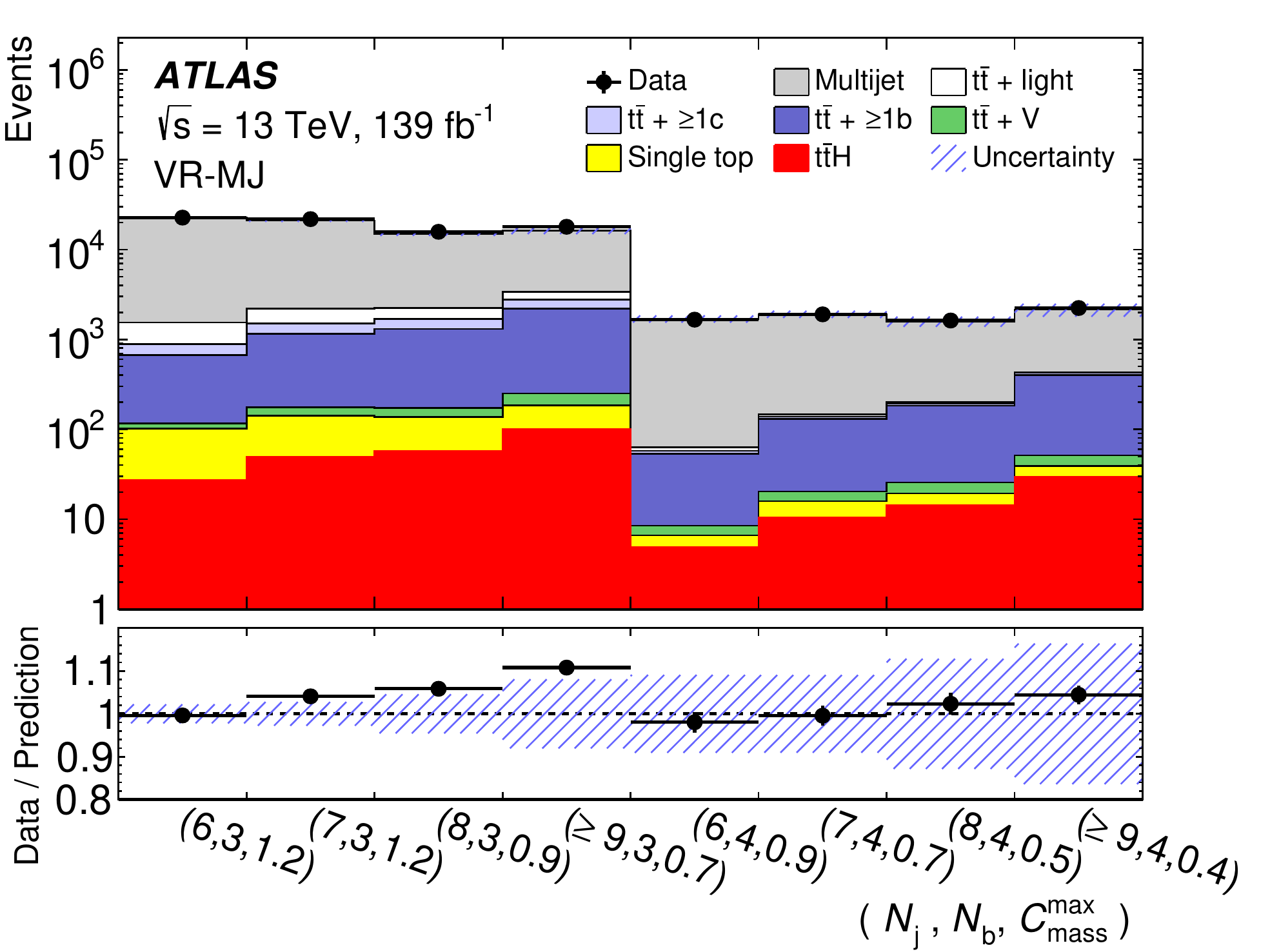}
\caption{Comparison between data and the predicted number of events with $N_{\textrm{j}}=$ 6, 7, 8 and $\geq 9$ and $N_{\textrm{b}}=$ 3 and 4 in the VR-MJ validation regions, which are based on a maximum value of the centrality mass,
$C_{\textrm{mass}}^{\textrm{max}}$. 
The bottom panel displays the ratios of data to the total prediction, uncertainty bars are statistical only. The systematic uncertainties listed in Section~\ref{sec:sys} are represented by the blue hatched area. }
\label{fig:cm}
\end{figure}

The \trfmj\ method is validated using two different comparisons with data: in the VR-MJ regions defined in Section~\ref{sec:analysis}, and in a separate set of  $Z$ + jets-enriched events. Figure~\ref{fig:cm} shows a comparison between measured and estimated event rates in VR-MJ. The data and predictions are in agreement within systematic uncertainties (described in Section~\ref{sec:sys}). 
 
\begin{figure}[h]
\centering
\includegraphics[width=0.75\textwidth]{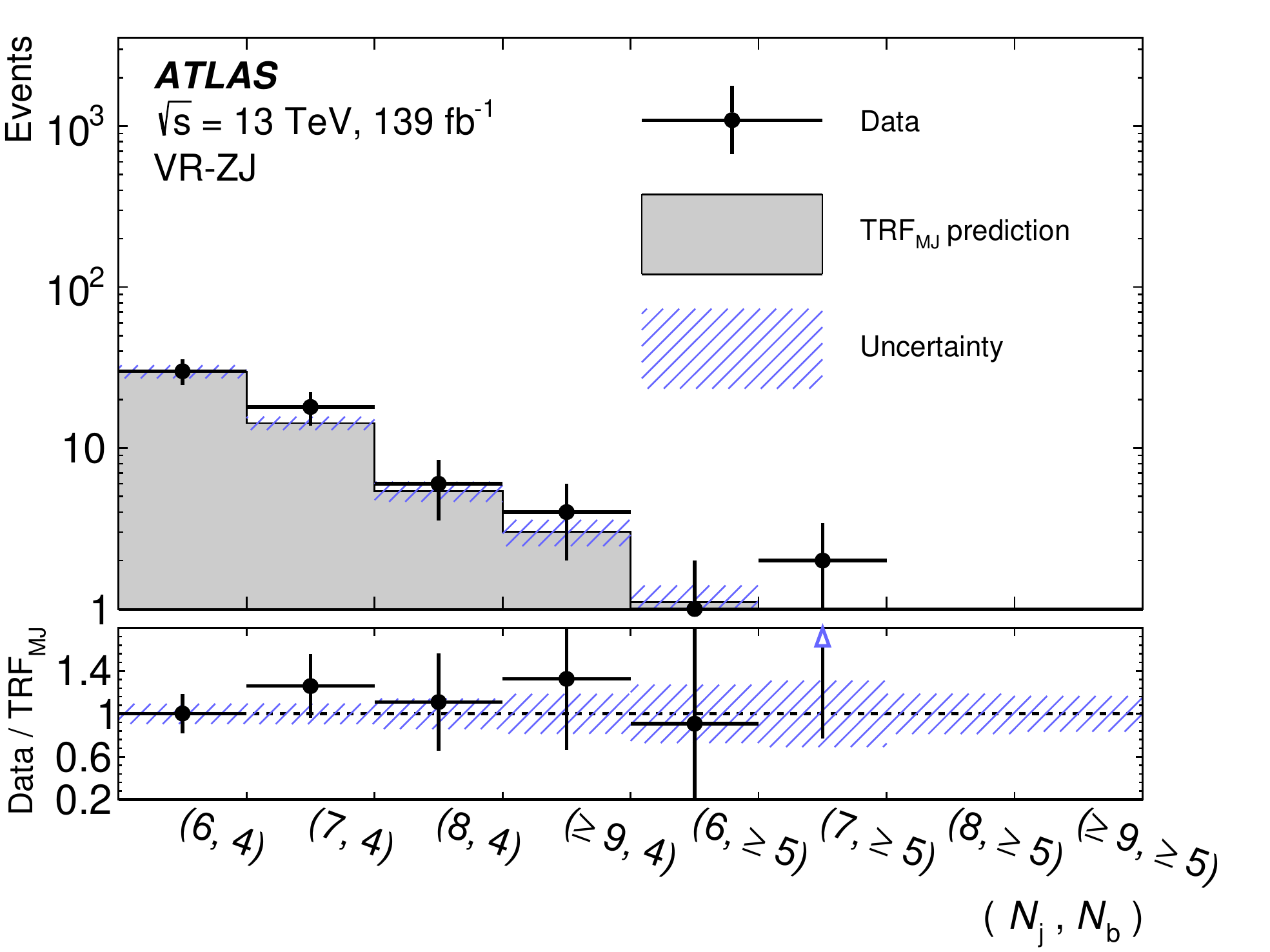}
\caption{Comparison between data and the number of events with $N_{\textrm{j}}=$ 6, 7, 8 and $\geq 9$ and $N_{\textrm{b}}=$4 and $\geq$5 predicted by the \trfmj\ method (grey histogram) in the VR-ZJ region, defined by the requirement of two isolated leptons with invariant mass larger than 60~\GeV. The bottom panel displays the ratios of data to the \trfmj\ prediction, uncertainty bars are statistical only. Systematic uncertainties in the \trfmj\ prediction are represented by the blue hatched area.}
\label{fig:valzjets}
\end{figure}

An independent test of the method is performed in $Z$ + jets-enriched events, referred as `VR-ZJ', where additional jets are produced by radiation and where \bbbar pairs arise from gluon splitting. In order to select events where a $Z$ boson decays into
pairs of
electrons or muons, events are required to pass a single-lepton trigger. Two opposite-sign, same-flavour, \textit{tight} electrons or muons are required to each have \pt $>$ 27~\GeV\ and a pair mass larger than 60~\GeV. Events are required to have at least five jets with \pt $>$ 25~\GeV\ and $|\eta|<$ 2.5, of which at least two must be $b$-tagged. The tagging probabilities \mjepstwo\ and \mjepsthree\ are derived from
five-jet VR-ZJ events and used to predict the number of events
with
$N_{\textrm{j}}=$ 6, 7, 8, $\geq9$ and $N_{\textrm{b}}=4$, $\geq5$.
As shown in Figure~\ref{fig:valzjets}, this statistically limited test further validates the \trfmj\ method.

\section{Systematic uncertainties}
\label{sec:sys}
Several sources of systematic uncertainty are considered that can affect the overall normalisation of signal and background samples and their relative contribution for different values of $N_{\textrm{j}}$ and $N_{\textrm{b}}$. In estimating the dominant multijet background from the data, systematic uncertainties arise from the assumptions made in obtaining the \trfmj\ background estimates. Uncertainties related to the theoretical modelling and due to the description of the detector response in simulated events are relevant only for the signal and background MC samples.
 
The main assumption of the \trfmj\ method is that it is possible
to define per-jet $b$-tagging probabilities (\mjepstwo\ and  \mjepsthree) in events with at least two or at least three $b$-tagged jets and, in particular, that
the variables used for the parameterisation are sensitive to the heavy-flavour composition of the jet sample. A second assumption is that the per-jet probabilities are independent of the jet multiplicity and, therefore, may be derived in a specific region, namely that with exactly five jets, and applied to regions with $N_{\textrm{j}}$ = 6, 7, 8 and $\geq$ 9 jets. The validity of these assumptions is verified using MC simulations. 
The \trfmj\ method is applied to \textsc{Pythia~8} MC dijet events, and the larger of a) the residual non-closure and b) the statistical uncertainty in the number of events with a given $b$-tagged jet multiplicity, is symmetrised and taken to be the systematic uncertainty associated with the method. Table~\ref{tab:TRFMJsys} shows the final \trfmj\ systematic uncertainty in the multijet background estimation in each ($N_{\textrm{j}}$, $N_{\textrm{b}}$) region. For $N_{\textrm{b}}$ = 4 the \trfmj\ uncertainties are dominated by the non-closure component, while for $N_{\textrm{b}}\geq$ 5, the statistical component dominates. The \trfmj\ uncertainties are the source of the largest systematic uncertainty for the analysis.
 
\begin{table}[hbt!]
\caption{Systematic uncertainties in the data-driven estimation of the multijet background using the \trfmj\ method.  The uncertainties are assessed using \textsc{Pythia~8} MC dijet events for each value of jet multiplicity ($N_{\textrm{j}}$) and $b$-tagged jet multiplicity ($N_{\textrm{b}}$) used in the final fit.}
\begin{center}
\begin{tabular}{  l  c  c  c }
 
\toprule
\multicolumn{2}{l}{\trfmj} & \multicolumn{2}{c}{$N_{\textrm{b}}$} \\
\multicolumn{2}{l}{uncertainty} & 4 & $\geq$5 \\
\midrule
\multirow{4}{*}{$N_{\textrm{j}}$} & 6 & ~~9\% & 27\% \\
& 7 & ~~9\% & 30\% \\
& 8 & 13\% & 18\% \\
& $\geq$9~~~ & 16\% & 14\% \\
\bottomrule
\end{tabular}\\
\end{center}
\label{tab:TRFMJsys}
\end{table}
 
The second largest contribution to the total systematic uncertainty arises from the modelling of the
$t\bar{t}$+jets background.
The
diagrams that contribute to \ttbin, \ttcin, and \ttbar+light production are different, and the
associated uncertainties may affect these processes differently
in different regions.
As a result, all uncertainties in $t\bar{t}$+jets background modelling, except the uncertainty in the inclusive cross-section, are considered to be uncorrelated among \ttbin, \ttcin, and \ttl.  
 
The uncertainty in the inclusive $t\bar{t}$ NNLO+NNLL
production cross-section is taken to be $\pm6\%$~\cite{Czakon:2011xx}.
This uncertainty includes effects from varying the factorisation and renormalisation scales, the PDF, $\alpha_{\textrm{S}}$, and the top-quark mass.
The normalisations of the \ttcin\ and \ttbin\ yields are taken from their fractional contribution to the nominal $t\bar{t}$+jets sample as generated using the \POWHEGBOX program. In addition to the uncertainty in the inclusive $t\bar{t}$ cross-section, an additional uncertainty of 50\%, based on the measurement
of the \ttbin\ and \ttcin\ normalisation factors
reported
in Ref.~\cite{Aaboud:2017rss}, is assigned  to the \ttcin\ and \ttbin\ production cross-sections.
 
The impact of the parton shower and hadronisation model uncertainties on the $t\bar{t}$+jets, $t\bar{t}H$ and $Wt$ single-top-quark yields is evaluated by comparing the sample from the nominal generator set-up with a sample produced with the NLO \POWHEGBOXV{v2}
generator using the NNPDF3.0 NLO PDF set. The latter events are interfaced with \HERWIGV{7.04}~\cite{Bahr:2008pv,Bellm:2015jjp}, using the H7UE set of tuned parameters~\cite{Bellm:2015jjp} and the MMHT2014LO PDF set \cite{Harland-Lang:2014zoa},
and processed using fast simulation of the detector response. The difference between the two predictions of the \ttbin\ event yield ranges from 20\% (33\%) for $N_{\textrm{j}}$ = 6 and $N_{\textrm{b}}$ = 4 ($\geq5$) to 46\% (60\%) in the region with $N_j\geq9$ and $N_{\textrm{b}}$ = 4 ($\geq5$).
 
To assess the uncertainty due to the choice of matching scheme, the \POWHEGBOX sample is compared with
a sample produced by \MGMCatNLO and \textsc{Pythia~8}. For the calculation of the hard scattering, \MGMCatNLOV{v2.6.0} with the
NNPDF3.0 NLO PDF set is used. The events are
processed
with \textsc{Pythia 8}.230,
using the A14 set of tuned parameters and the NNPDF2.3 LO PDF set, and
the
fast simulation of the detector response. The uncertainty, which is obtained from the difference in yield between the two models and is symmetrised, affects both the normalisation and the $N_{\textrm{j}}$- and $N_{\textrm{b}}$-dependence of background rates. It is largest for large values of the jet and  $b$-tagged jet multiplicities. For \ttbin, it reaches 25\% for $N_{\textrm{j}}$ = 8, $\geq9$ and $N_{\textrm{b}}$ = 4, and 41\% (32\%)
for $N_{\textrm{j}}$ = 8 ($\geq9$) and $N_{\textrm{b}}\geq5$.
 
The effect of renormalisation and factorisation scale uncertainties and PDF uncertainties is evaluated for \ttH\ and \ttV\ events. For the former, the scales are varied simultaneously by common factors of 2.0 and 0.5. For the latter, the envelope of the 100 variations for NNPDF3.0 NLO~\cite{Ball:2014uwa} are taken into account.
An uncertainty of $\pm 5\%$ is assigned to the total cross-section for single-top production~\cite{Kidonakis:2011wy,Kidonakis:2010ux,Kidonakis:2010tc}. For both the \ttH\ and single-top events, additional uncertainties due to initial- and final-state radiation and the choice of generator
are evaluated in a manner similar to that used for \ttbar + jets. The uncertainty in the amount of interference between $Wt$ and \ttbar\
production at NLO is assessed by comparing samples using the default `diagram removal' scheme with those using an alternative `diagram subtraction' scheme~\cite{Frixione:2008yi}.
All modelling uncertainties from non-$t\bar{t}$+jets simulated backgrounds are, after investigation, found to be negligible.

The uncertainties assigned to the expected signal yield for the SUSY benchmark processes considered include the experimental uncertainties related to the luminosity and to the detector modelling, which are dominated by the modelling of the jet energy scale and the $b$-tagging efficiencies.
For example, for the $\tilde{t} \rightarrow b \tilde{\chi}^{+}_{1}$($\tilde{\chi}^{+}_{1}\rightarrow \bar{b} \bar{b} \bar{s}$ and c.c.) signal model, the $b$-tagging uncertainties in the region $N_j\geq9$ and $N_{\textrm{b}}$ = 4 are approximatively 10\%, and the jet-related uncertainties of the signal yields are in the range of 3--5\%. The uncertainties in the signal yields related to the modelling of additional jet radiation are studied by varying the factorisation, renormalisation, and jet-matching scales as well as the parton-shower tune in the simulation. The corresponding uncertainties are small for most of the signal parameter space and are largest for small top-squark masses, where they reach 7\%. The uncertainty in the signal cross-section ranges between 8\% and 11\% for a top-squark mass in the range 600--1000~\GeV.

\section{Results}
\label{sec:fit}
 
The events are allocated to ($N_{\textrm{j}},N_{\textrm{b}}$) regions with different signal-to-background ratios in order to constrain systematic
uncertainties and
to improve the
separation of signal and background.
Then, in each region, the total signal and background yields, shown in Tables~\ref{tab:prefit4} and~\ref{tab:prefit5}, are used in combination as the input for the statistical analysis to extract the final results.
 
\begin{table}[htb!]
\begin{center}
\caption{Event yields from background predictions and data in the regions with $N_{\textrm{j}}=$ 6, 7, 8 or $\geq9$ and $N_{\textrm{b}}=4$. The quoted uncertainties are the sum in quadrature of the statistical and systematic uncertainties in the yields for all samples. The individual background uncertainties can be larger than the total uncertainty due to correlations between parameters. }
\begin{tabular}{l |
>{\collectcell\num}r<{\endcollectcell}
@{${}\pm{}$}
>{\collectcell\num}l<{\endcollectcell}
>{\collectcell\num}r<{\endcollectcell}
@{${}\pm{}$}
>{\collectcell\num}l<{\endcollectcell}
>{\collectcell\num}r<{\endcollectcell}
@{${}\pm{}$}
>{\collectcell\num}l<{\endcollectcell}
>{\collectcell\num}r<{\endcollectcell}
@{${}\pm{}$}
>{\collectcell\num}l<{\endcollectcell}
}
 
\toprule
& \multicolumn{8}{c}{($N_{\textrm{j}}$, $N_{\textrm{b}}$)} \\
Process & \multicolumn{2}{ c}{(6, 4)} & \multicolumn{2}{c}{(7, 4)} & \multicolumn{2}{c}{(8, 4)} & \multicolumn{2}{c}{($\geq$9, 4)} \\
\midrule
Multijet   & 1760 & 170 & 1920 & 180 & 1510 & 210 & 1870 & 350 \\
$t\bar{t}$ + light   & 6 & 4 & 8.0 & 3.4 & 6 & 4 & 8 & 7 \\
$t\bar{t}$ + $\geq$1$c$   & 4.1 & 2.9 & 8 & 5 & 11 & 6 & 22 & 17 \\
$t\bar{t}$ + $\geq$1$b$   & 45 & 26 & 110 & 70 & 160 & 100 & 350 & 260 \\
$t\bar{t}$ + $W$   & 0.055 & 0.032 & 0.26 & 0.07 & 0.30 & 0.10 & 1.34 & 0.28 \\
$t\bar{t}$ + $Z$   & 1.8 & 0.4 & 4.3 & 1.0 & 6.0 & 1.5 & 10.9 & 2.3 \\
$Wt$   & 1.7 & 2.0 & 5 & 5 & 5.1 & 3.1 & 10 & 11 \\
$t\bar{t}H$   & 4.9 & 0.9 & 10.5 & 1.7 & 14.2 & 2.4 & 29 & 8 \\
\midrule
Total background & 1820 & 170 & 2060 & 190 & 1710 & 220 & 2300 & 400 \\
Data   & \multicolumn{2}{ c}{1660} & \multicolumn{2}{c}{1901} & \multicolumn{2}{c}{1624} & \multicolumn{2}{c}{2237} \\
\bottomrule
\end{tabular}
\label{tab:prefit4}
\end{center}
\end{table}

\begin{table}[htbp]
\begin{center}
\caption{Event yields from background predictions and data in the regions with $N_{\textrm{j}}=$ 6, 7, 8 or $\geq9$ and $N_{\textrm{b}}\geq$ 5. The quoted uncertainties are the sum in quadrature of the statistical and systematic uncertainties in the yields for all samples. The individual background uncertainties can be larger than the total uncertainty due to correlations between parameters. }
\begin{tabular}{ l |
>{\collectcell\num}r<{\endcollectcell}
@{${}\pm{}$}
>{\collectcell\num}l<{\endcollectcell}
>{\collectcell\num}r<{\endcollectcell}
@{${}\pm{}$}
>{\collectcell\num}l<{\endcollectcell}
>{\collectcell\num}r<{\endcollectcell}
@{${}\pm{}$}
>{\collectcell\num}l<{\endcollectcell}
>{\collectcell\num}r<{\endcollectcell}
@{${}\pm{}$}
>{\collectcell\num}l<{\endcollectcell}
}
\toprule
& \multicolumn{8}{c}{($N_{\textrm{j}}$, $N_{\textrm{b}}$)} \\  [0.15cm]
Process & \multicolumn{2}{c}{(6, $\geq$5)} & \multicolumn{2}{c}{(7, $\geq$5)} & \multicolumn{2}{c}{(8, $\geq$5)} & \multicolumn{2}{c}{($\geq$9, $\geq$5)} \\
\midrule
Multijet  & 49 & 13 & 75 & 23 & 74 & 14 & 123 & 20 \\
$t\bar{t}$ + light   & \multicolumn{2}{c}{$<$0.01} & 0.3 & 0.6 & \multicolumn{2}{c}{$<$0.01} & 0.00 & 0.04 \\
$t\bar{t}$ + $\geq$1$c$   & \multicolumn{2}{c}{$<$0.01} & 0.016 & 0.029 & 0.3 & 0.4 & 0.26 & 0.31 \\
$t\bar{t}$ + $\geq$1$b$   & 1.2 & 0.9 & 3.9 & 2.7 & 7 & 6 & 28 & 25 \\
$t\bar{t}$ + $W$   & \multicolumn{2}{c}{$<$0.01} & 0.005 & 0.007 & 0.021 & 0.025 & 0.090 & 0.035 \\
$t\bar{t}$ + $Z$   & 0.05 & 0.05 & 0.22 & 0.12 & 0.7 & 0.4 & 0.7 & 0.7 \\
$Wt$   & \multicolumn{2}{c}{$<$0.01} & \multicolumn{2}{c}{$<$0.01} & 0.00 & 0.13 & 0.9 & 1.2 \\
$t\bar{t}H$   & 0.12 & 0.05 & 0.49 & 0.13 & 0.82 & 0.21 & 2.9 & 1.5 \\
\midrule
Total background & 50 & 13 & 80 & 23 & 84 & 15 & 156 & 27 \\
Data   & \multicolumn{2}{c}{35} & \multicolumn{2}{c}{75} & \multicolumn{2}{c}{80} & \multicolumn{2}{c}{179} \\
\bottomrule
\end{tabular}
\label{tab:prefit5}
\end{center}
\end{table}

Hypothesis testing is performed using a modified frequentist method as implemented in \textsc{ RooStats}~\cite{RooFitManual} and is based on a
profile likelihood which takes into account the systematic uncertainties as nuisance parameters. This procedure minimises the impact of systematic uncertainties on the search sensitivity by taking advantage of the highly populated, background-dominated ($N_\textrm{j},N_\textrm{b}$) regions included in the likelihood fit. The signal-strength parameter, $\mu_{\tilde{t}\tilde{t}^{*}}$, defined for positive values and corresponding to the signal normalisation,
is unconstrained
in the profile-likelihood fit. The normalisation of each component of the background and $\mu_{\tilde{t}\tilde{t}^{*}}$ are determined simultaneously from the fit to the data.
 
Individual sources of systematic uncertainty are taken as uncorrelated. Contributions from
$t\bar{t}$ + $\geq1b$, $t\bar{t}$ + $\geq1c$, $t\bar{t}$ + light, $t\bar{t}$ + $V$, $t\bar{t}H$ and
single-top-quark backgrounds are constrained by the uncertainties of the respective theoretical calculations, the uncertainty in the luminosity (described in Section~\ref{sec:sim}), and experimental data. The \trfmj\ uncertainty is taken as uncorrelated across regions because of its large statistical component. In all cases, the profile-likelihood-ratio test is used to establish 95\% confidence intervals using the CL$_\textrm{s}$~\cite{Read:2002hq} prescription. The likelihood is configured differently for the model-independent and model-dependent hypothesis tests.
 
For the model-independent test, a profile-likelihood fit is performed independently in the two SR$_{\textrm{discovery}}$ regions with ($N_{\textrm{j}} \geq$ 8, $N_{\textrm{b}} \geq$ 5) and ($N_{\textrm{j}} \geq$ 9, $N_{\textrm{b}} \geq$ 5). This test is used to search for, and to compute generic exclusion limits on, the potential contribution from a hypothetical BSM signal in the given SR$_{\textrm{discovery}}$ regions.
 
For the model-dependent test, assuming a specific top-squark model with variable mass values,
tests of the signal-plus-background hypothesis, i.e.\ $\mu_{\tilde{t}\tilde{t}^{*}}$ = 1, are formed for a series of values of $m_{\tilde{t}}$ and $m_{\tilde{\chi}^0_{1,2},\tilde{\chi}^{\pm}_1}$. These are used to derive exclusion limits for the specific top-squark model.
The full set of regions,
$N_{\textrm{j}}=$ 6, 7, 8 and $\geq$ 9 and $N_{\textrm{b}}=$ 4 and $\geq$ 5,
is employed in the likelihood. The expected signal contribution, as predicted by the given model, is considered in all regions and is scaled by $\mu_{\tilde{t}\tilde{t}^{*}}$.
 
\begin{figure}[htb]
\centering
\includegraphics[width=0.7\textwidth]{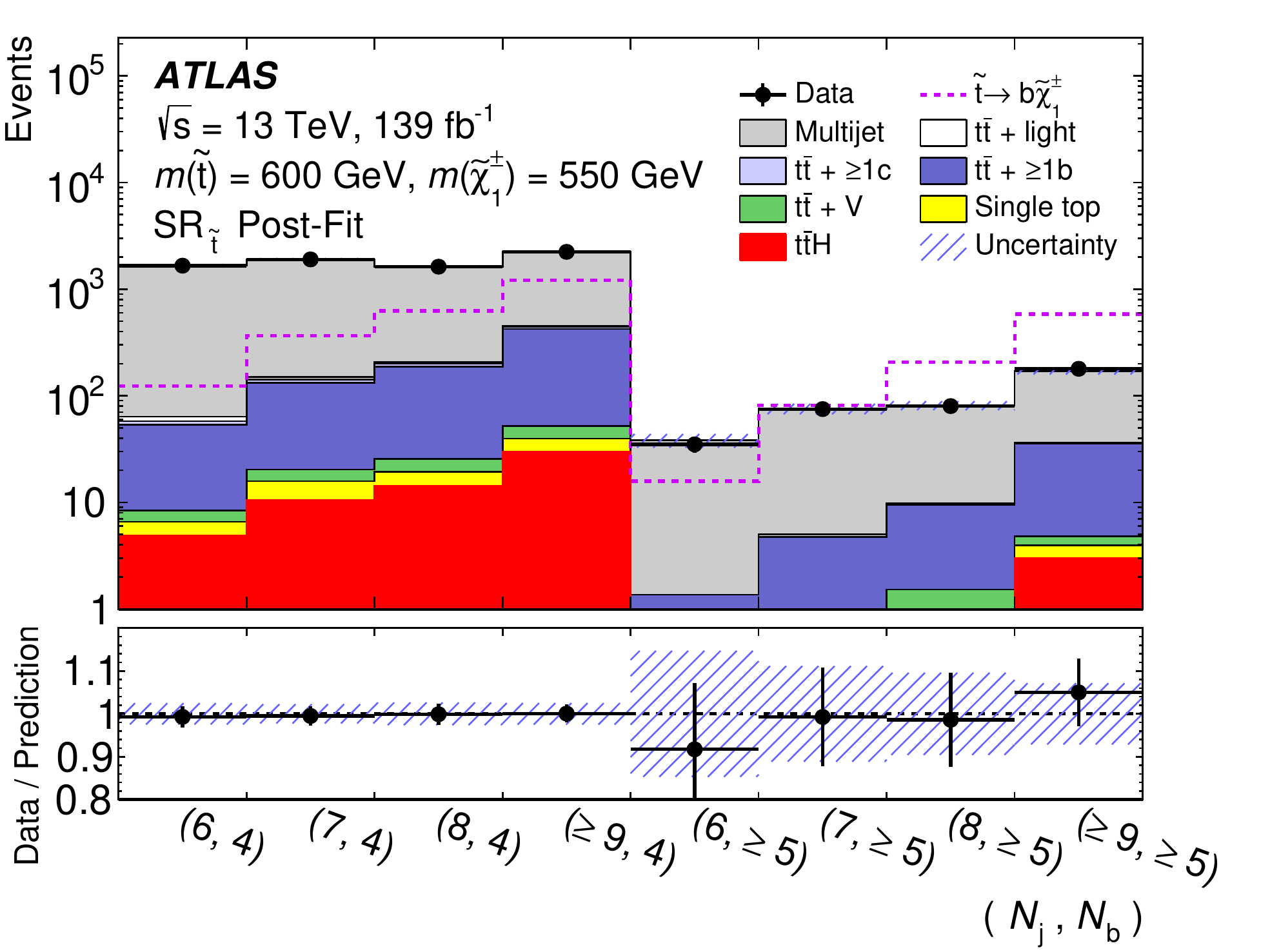}
\caption{ Expected background and observed number of events in different jet and $b$-tag multiplicity bins. The background is estimated by including all bins in a background-only fit and is plotted separately for each contribution. An example signal yield for $ \tilde{t} \rightarrow \bar{b} \tilde{\chi}^{+}_{1}$  ($\tilde{\chi}^{+}_{1} \rightarrow \bar{b} \bar{b} \bar{s}$ and c.c.) production with $m_{\tilde{t}}$ = 600~\GeV\ and $m_{\tilde{\chi}^{\pm}_{1}}$ = 550~\GeV\ is overlaid. The bottom panel displays the ratios of data to the total prediction, uncertainty bars are statistical only. All uncertainties, which can be correlated across bins, are included in the error bands (hatched regions). \label{fig:pf} }
\end{figure}

Figure~\ref{fig:pf} shows the observed numbers of data events compared with the fitted background model. The likelihood fit is configured using the model-dependent set-up where all bins are input to the fit, and $\mu_{\tilde{t}\tilde{t}^{*}}$ is set to zero. This configuration is also referred to as the \textit{background-only} fit
and includes no free-floating
parameters, only nuisance parameters with Gaussian constraints. An example signal model is also shown in the figure to illustrate the
separation
between the signal and the background. 
 
\subsection{Model-independent interpretation}
 
The model-independent results are calculated from the observed number of events and the background predictions in the two SR$_{\textrm{discovery}}$ regions. The observed number of events and the backgrounds obtained from the fits are shown for both SR$_{\textrm{discovery}}$ regions in Table~\ref{tab:modind}.
 
\begin{table}[tbp]
\caption{Fitted background yields in ($N_{\textrm{j}} \geq$ 8, $N_{\textrm{b}} \geq$ 5) and ($N_{\textrm{j}} \geq$ 9, $N_{\textrm{b}} \geq$ 5)
signal regions. The individual background uncertainties can be larger
than the total uncertainty due to correlations between parameters.  }
\begin{center}
\begin{tabular}{ l
>{\collectcell\num}r<{\endcollectcell}
@{${}\pm{}$}
>{\collectcell\num}l<{\endcollectcell}
>{\collectcell\num}r<{\endcollectcell}
@{${}\pm{}$}
>{\collectcell\num}l<{\endcollectcell}
}
\toprule
Process & \multicolumn{2}{c}{$N_{\textrm{j}} \geq$8, $N_{\textrm{b}} \geq$5} & \multicolumn{2}{c}{$N_{\textrm{j}} \geq$9, $N_{\textrm{b}} \geq$5} \\
\midrule
Multijet  & 200 & 40 & 123 & 20 \\
$t\bar{t}$ + $\geq$1$c$   &  0.6 & 0.6 & 0.29 & 0.33 \\
$t\bar{t}$ + $\geq$1$b$   & 26 & 20 & 20 & 15 \\
$t\bar{t}$ + $W$   & 0.11 & 0.05 & 0.09 & 0.04 \\
$t\bar{t}$ + $Z$   & 1.4 & 0.7 & 0.8 & 0.7 \\
$Wt$  & 0.9 & 0.8 & 0.9 & 1.2 \\
$t\bar{t}H$   & 3.7 & 1.6 & 2.9 & 1.4 \\
\midrule
Total background & 230 & 40 & 147 & 20 \\
Data   & \multicolumn{2}{c}{259} & \multicolumn{2}{c}{179} \\
\bottomrule
\end{tabular}
\end{center}
\label{tab:modind}
\end{table}
 
Model-independent 95\% CL upper limits on the expected and observed number of BSM events, $N^{95}_{\text{exp}}$ and $\sigma^{95}_{\text{obs}}$, that may contribute to the signal regions are computed from the observed number of events and the fitted background.
Normalising these results by the integrated luminosity, $L$, of the data sample, allows them to be interpreted as upper limits on the visible BSM cross-section  $\sigma^{95}_{\text{obs}}$, defined as:
\begin{equation}\nonumber
\sigma^{95}_{\text{obs}} = \sigma_{\textrm{prod}} \times \textrm{A} \times \epsilon = \dfrac{N^{95}_{\text{obs}}}{L},
\end{equation}
where $\sigma_{\textrm{prod}}$ is the production cross-section. The resulting limits are presented in Table~\ref{tab:n}. In addition, the $p_0$ values, which quantify the probability that a background-only hypothesis results in a fluctuation giving an event yield equal to or larger than the one observed in the data, are calculated, as are the corresponding Gaussian significance values $Z$.

\begin{table}[htb]
\caption{Observed 95\% CL model-independent upper limits on
the visible BSM cross-section, $\sigma^{95}_{\text{obs}}$, expressed in fb,
along with the observed (expected) limits, $N^{95}_{\text{obs}}$ ($N^{95}_{\text{exp}}$),
on the number of excess events.  The limits are determined for two signal regions,
($N_{\textrm{j}} \geq$ 8, $N_{\textrm{b}} \geq$ 5) and ($N_{\textrm{j}} \geq$ 9, $N_{\textrm{b}} \geq$ 5).  The $p_0$ value quantifies the probability that the background-only hypothesis would result in a fluctuation that gives an event yield equal to or larger than the one observed in the data, and $Z$
is the corresponding Gaussian
significance.}
\begin{center}
\begin{tabular}{lSlSS}
\toprule
Signal region & {$\sigma^{95}_{\text{obs}}$ [fb]} & \multicolumn{1}{c}{$N^{95}_{\text{obs}}$}
& \multicolumn{1}{c}{$N^{95}_{\text{exp}}$} & \multicolumn{1}{c}{$p_0$ ($Z$)}\\
\midrule
$N_{\textrm{j}}\geq$8, $N_{\textrm{b}}\geq$5               &  0.76           & 105        &  {$85^{+30}_{-24}$} & {0.24 (0.7)} \\ [0.15cm]
$N_{\textrm{j}}\geq$9, $N_{\textrm{b}}\geq$5  &  0.54          & ~~75         &   {$52^{+20}_{-15}$}  &  {0.11 (1.2)} \\
\bottomrule
\end{tabular}
\end{center}
\label{tab:n}
\end{table}

\subsection{Model-dependent interpretation}
 
\begin{figure}[htb]
\centering
\includegraphics[width=0.7\textwidth]{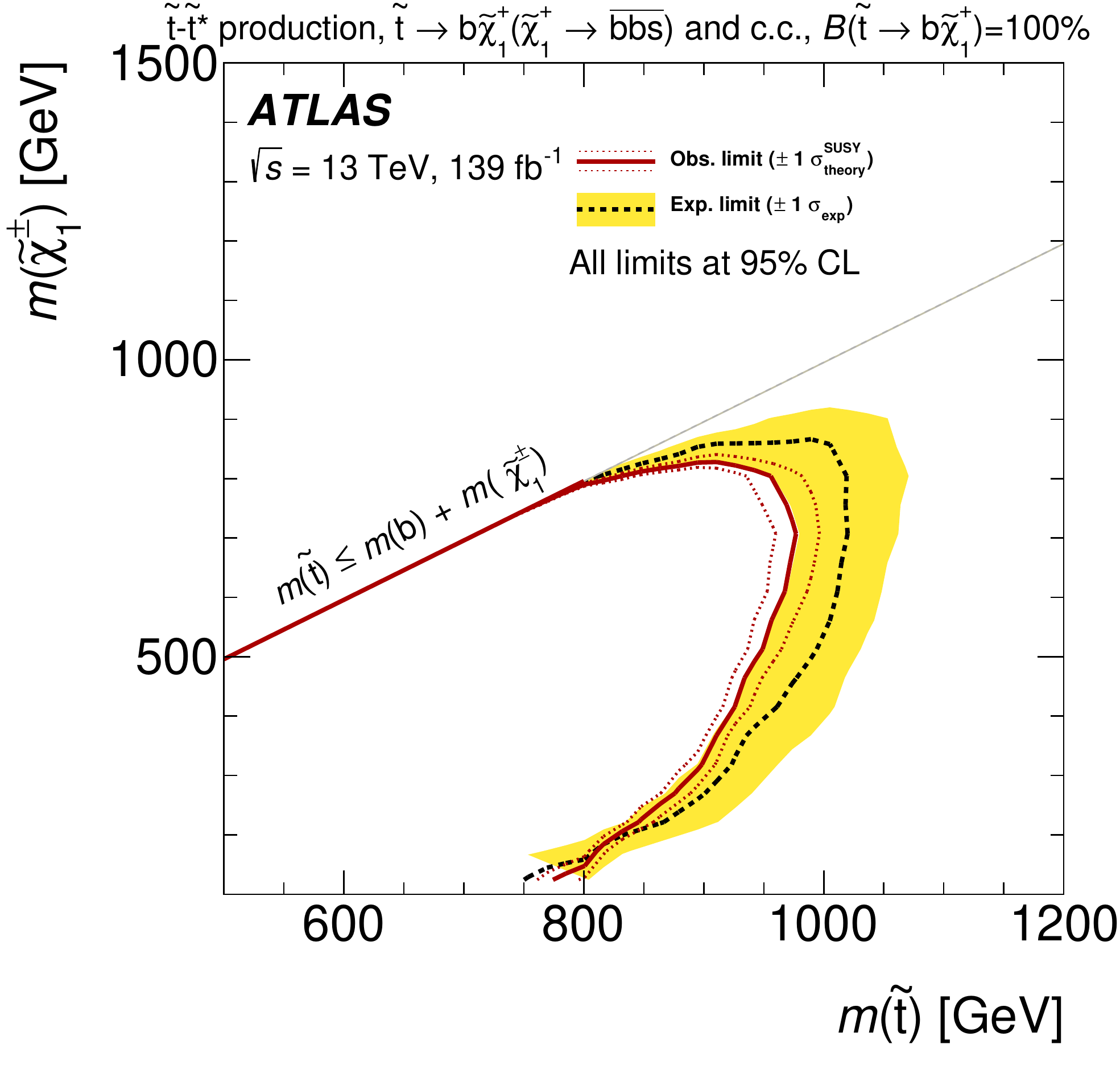}
\caption{ Observed and expected exclusion contours for the $\tilde{t}$ and $\chi^{\pm}_{1}$ masses
in a
top-squark production
model with RPV decays of the  $\chi^{\pm}_{1}$. Limits are shown
for $B(\tilde{t} \rightarrow b \chi^{+}_{1})$ equal to unity. The contours of the band around the expected limit are the $\pm 1 \sigma$ variations, including all uncertainties except theoretical uncertainties in the signal cross-section.
The dotted lines around the observed limit illustrate the change in the observed limit as the nominal signal cross-section is scaled up
and down by the theoretical uncertainty. All limits are computed at 95\% CL.
The results are constrained by the kinematic limits of the top-squark decay into a chargino and a bottom quark (diagonal line).
\label{fig:LC} }
\end{figure}

For each signal model probed, the fit is configured using the model-dependent set-up, as detailed in the first part of Section~\ref{sec:fit}. Figure~\ref{fig:LC} shows exclusion limits at the 95\% confidence level in the top-squark production model when $B(\tilde{t} \rightarrow b \tilde{\chi}^{+}_{1})$ is assumed to be unity. For this model, top-squark masses are excluded up to 950~\GeV\ for chargino masses close to the kinematic threshold for producing this final state. For lower values of the chargino mass, the limit weakens such that for chargino masses of around 200~\GeV, the top-squark mass is constrained to be more than 800~\GeV. In this phase space region, the signal is concentrated at lower $N_{\textrm{j}}$ and $N_{\textrm{b}}$ values where the background is larger.
 
\begin{figure}[htb]
\centering
\includegraphics[width=0.7\textwidth]{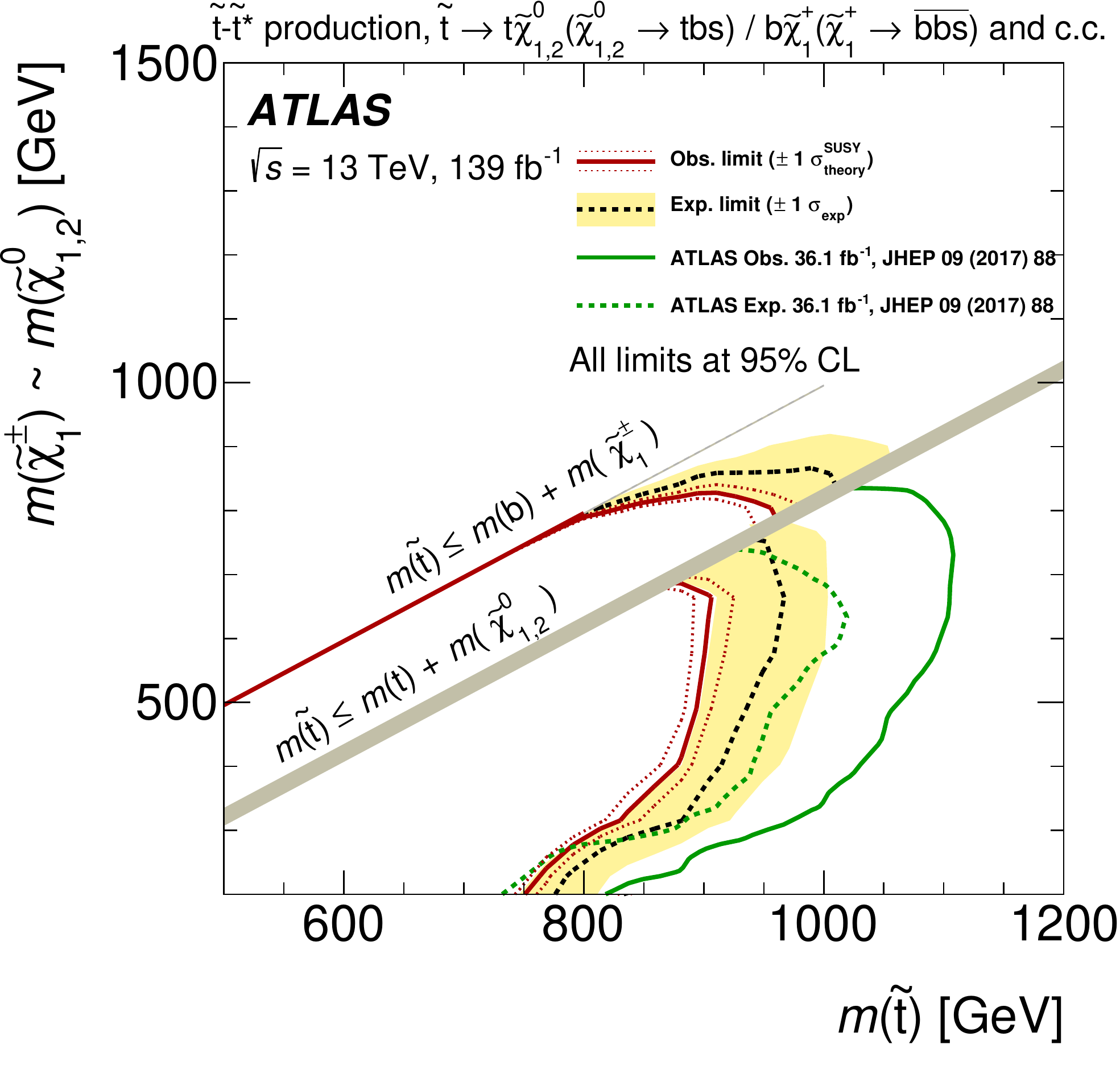}
\caption{ Observed and expected exclusion contours for the $\tilde{t}$ and $\chi^{\pm}_{1}$ masses
in a
top-squark production
model with RPV decays of the  $\chi^{\pm}_{1}$ and of the $\tilde{\chi}^0_{1,2}$. Limits are shown
in the case of a higgsino LSP. The contours of the band around
the expected limit are the $\pm 1 \sigma$ variations, including all uncertainties except theoretical uncertainties in the signal cross-section.
The dotted lines around the observed limit illustrate the change in the observed limit as the nominal signal cross-section is scaled up
and down by the theoretical uncertainty.
All limits are computed at 95\% CL.
The results are constrained by the kinematic limits of the top-squark decay into a chargino and a bottom quark (upper diagonal line) and into a neutralino and a top quark (lower diagonal line), respectively.
Also shown are the limits from Ref.~\cite{Aaboud:2017faq}.
\label{fig:LH} }
\end{figure}

The limits for higgsino LSPs are shown in Figure~\ref{fig:LH}. In the region $m_{\tilde{t}} - m_{\tilde{\chi}^0_{1,2}, \tilde{\chi}^\pm_{1}} \geq m_\text{top}$ the sensitivity of the analysis is lower than in the pure $\tilde{t} \rightarrow b \tilde{\chi}^{\pm}_{1}$ case because contributions to the signal that have one leptonically decaying top quark fail the lepton-veto requirement. The large contribution of the multijet background reduces the present sensitivity relative to a previous ATLAS search that analysed events characterised by the presence of a lepton plus jets~\cite{Aaboud:2017faq}.
 
\FloatBarrier
 
\FloatBarrier
\section{Conclusion}
\label{sec:conclusion}
 
A search for physics beyond the Standard Model in events with high jet multiplicity and a large number of $b$-tagged jets is described in this
paper.
The search uses 139~\ifb\ of $\sqrt{s} = 13~\TeV$ proton--proton collision data collected by the ATLAS experiment at the LHC.
In contrast to
many previous searches in similar final states, leptons are vetoed and no requirement
is placed on the missing transverse momentum in the
event.
The multijet background dominates the observed yields and is estimated using a data-driven technique based on an extrapolation from events with low $b$-jet multiplicity to the high $b$-jet multiplicities.
No significant excess over the SM expectation is observed,
and model-independent limits on the contribution of new phenomena to the signal-region yields are computed. In the context of
a
model with direct top-squark production and RPV decays of the higgsinos, the data exclude top squarks with masses up to 950~\GeV\ in the region $m_{\tilde{t}} - m_{\tilde{\chi}^0_{1,2},\tilde{\chi}^\pm_{1}} \leq \mtop$, where this analysis has exclusive sensitivity. The results represent the first limits from ATLAS on the production of top squarks that decay exclusively into a chargino and a $b$-quark.

\section*{Acknowledgements}
 

We would like to thank Gilbert Moultaka for helpful discussions on the phenomenological aspects of the analysis. We thank CERN for the very successful operation of the LHC, as well as the
support staff from our institutions without whom ATLAS could not be
operated efficiently.
 
We acknowledge the support of ANPCyT, Argentina; YerPhI, Armenia; ARC, Australia; BMWFW and FWF, Austria; ANAS, Azerbaijan; SSTC, Belarus; CNPq and FAPESP, Brazil; NSERC, NRC and CFI, Canada; CERN; ANID, Chile; CAS, MOST and NSFC, China; COLCIENCIAS, Colombia; MSMT CR, MPO CR and VSC CR, Czech Republic; DNRF and DNSRC, Denmark; IN2P3-CNRS and CEA-DRF/IRFU, France; SRNSFG, Georgia; BMBF, HGF and MPG, Germany; GSRT, Greece; RGC and Hong Kong SAR, China; ISF and Benoziyo Center, Israel; INFN, Italy; MEXT and JSPS, Japan; CNRST, Morocco; NWO, Netherlands; RCN, Norway; MNiSW and NCN, Poland; FCT, Portugal; MNE/IFA, Romania; MES of Russia and NRC KI, Russia Federation; JINR; MESTD, Serbia; MSSR, Slovakia; ARRS and MIZ\v{S}, Slovenia; DST/NRF, South Africa; MICINN, Spain; SRC and Wallenberg Foundation, Sweden; SERI, SNSF and Cantons of Bern and Geneva, Switzerland; MOST, Taiwan; TAEK, Turkey; STFC, United Kingdom; DOE and NSF, United States of America. In addition, individual groups and members have received support from BCKDF, CANARIE, Compute Canada and CRC, Canada; ERC, ERDF, Horizon 2020, Marie Sk{\l}odowska-Curie Actions and COST, European Union; Investissements d'Avenir Labex, Investissements d'Avenir Idex and ANR, France; DFG and AvH Foundation, Germany; Herakleitos, Thales and Aristeia programmes co-financed by EU-ESF and the Greek NSRF, Greece; BSF-NSF and GIF, Israel; La Caixa Banking Foundation, CERCA Programme Generalitat de Catalunya and PROMETEO and GenT Programmes Generalitat Valenciana, Spain; G\"{o}ran Gustafssons Stiftelse, Sweden; The Royal Society and Leverhulme Trust, United Kingdom.
 
The crucial computing support from all WLCG partners is acknowledged gratefully, in particular from CERN, the ATLAS Tier-1 facilities at TRIUMF (Canada), NDGF (Denmark, Norway, Sweden), CC-IN2P3 (France), KIT/GridKA (Germany), INFN-CNAF (Italy), NL-T1 (Netherlands), PIC (Spain), ASGC (Taiwan), RAL (UK) and BNL (USA), the Tier-2 facilities worldwide and large non-WLCG resource providers. Major contributors of computing resources are listed in Ref.~\cite{ATL-SOFT-PUB-2020-001}.
 

\clearpage

\printbibliography

\clearpage 
 
\begin{flushleft}
\hypersetup{urlcolor=black}
{\Large The ATLAS Collaboration}

\bigskip

\AtlasOrcid[0000-0002-6665-4934]{G.~Aad}$^\textrm{\scriptsize 102}$,    
\AtlasOrcid[0000-0002-5888-2734]{B.~Abbott}$^\textrm{\scriptsize 128}$,    
\AtlasOrcid{D.C.~Abbott}$^\textrm{\scriptsize 103}$,    
\AtlasOrcid[0000-0002-2788-3822]{A.~Abed~Abud}$^\textrm{\scriptsize 36}$,    
\AtlasOrcid[0000-0002-1002-1652]{K.~Abeling}$^\textrm{\scriptsize 53}$,    
\AtlasOrcid[0000-0002-2987-4006]{D.K.~Abhayasinghe}$^\textrm{\scriptsize 94}$,    
\AtlasOrcid[0000-0002-8496-9294]{S.H.~Abidi}$^\textrm{\scriptsize 166}$,    
\AtlasOrcid[0000-0002-8279-9324]{O.S.~AbouZeid}$^\textrm{\scriptsize 40}$,    
\AtlasOrcid{N.L.~Abraham}$^\textrm{\scriptsize 155}$,    
\AtlasOrcid[0000-0001-5329-6640]{H.~Abramowicz}$^\textrm{\scriptsize 160}$
% ...
\bigskip
\\

% ...
$^{36}$CERN, Geneva; Switzerland.\\
% ...
$^{40}$Niels Bohr Institute, University of Copenhagen, Copenhagen; Denmark.\\
% ...
$^{53}$II. Physikalisches Institut, Georg-August-Universit\"{a}t G\"ottingen, G\"ottingen; Germany.\\
% ...
$^{94}$Department of Physics, Royal Holloway University of London, Egham; United Kingdom.\\
% ...
$^{102}$CPPM, Aix-Marseille Universit\'e, CNRS/IN2P3, Marseille; France.\\
$^{103}$Department of Physics, University of Massachusetts, Amherst MA; United States of America.\\
% ...
$^{128}$Homer L. Dodge Department of Physics and Astronomy, University of Oklahoma, Norman OK; United States of America.\\
% ...
$^{155}$Department of Physics and Astronomy, University of Sussex, Brighton; United Kingdom.\\
% ...
$^{160}$Raymond and Beverly Sackler School of Physics and Astronomy, Tel Aviv University, Tel Aviv; Israel.\\
% ...
$^{166}$Department of Physics, University of Toronto, Toronto ON; Canada.\\
% ...

\end{flushleft}


\end{document}